\documentclass[12pt,a4paper]{article}
\pdfoutput=1
\usepackage[utf8]{inputenc}
\usepackage[english]{babel}
\usepackage{amsmath}
\usepackage{subfigure}
\usepackage{amsfonts}
\usepackage{amssymb}
\usepackage{graphicx}
\newcommand{\be}{\begin{equation}}
\newcommand{\ee}{\end{equation}}
\newcommand{\bea}{\begin{eqnarray}}
\newcommand{\eea}{\end{eqnarray}}
\newcommand{\lt}{\left}
\newcommand{\rt}{\right}

\catcode`@=12
\def\la{\mathrel{\mathchoice {\vcenter{\offinterlineskip\halign{\hfil
$\displaystyle##$\hfil\cr<\cr\sim\cr}}}
{\vcenter{\offinterlineskip\halign{\hfil$\textstyle##$\hfil\cr<\cr\sim\cr}}}
{\vcenter{\offinterlineskip\halign{\hfil$\scriptstyle##$\hfil\cr<\cr\sim\cr}}}
{\vcenter{\offinterlineskip\halign{\hfil$\scriptscriptstyle##$\hfil\cr<\cr\sim
\cr}}}}}

\begin{document}
\thispagestyle{empty}
\begin{center}
{\Large\bf 
{Two component WIMP-FImP dark matter model with singlet fermion, scalar and 
pseudo scalar}}\\
\vspace{1cm}
{{\bf Amit Dutta Banik} \footnote{email: amit.duttabanik@saha.ac.in},
{\bf Madhurima Pandey} \footnote{email: madhurima.pandey@saha.ac.in} 
{\bf Debasish Majumdar} \footnote{email: debasish.majumdar@saha.ac.in}}\\
{\normalsize \it Astroparticle Physics and Cosmology Division,}
{\normalsize \it Saha Institute of Nuclear Physics, HBNI} \\
{\normalsize \it 1/AF Bidhannagar, Kolkata 700064, India}\\
\vspace{0.25cm}
{\bf Anirban Biswas} \footnote{email: anirbanbiswas@hri.res.in}\\
{\normalsize \it Harish Chandra Research Institute}
{\normalsize \it Chhatnag Road, Jhusi, Allahabad, India}\\
\vspace{1cm}
\end{center}
\begin{abstract}
We explore a two component dark matter model with  
a fermion and a scalar. In this scenario the  
Standard Model (SM) is extended by a fermion, a scalar and an 
additional pseudo scalar. The fermionic component is assumed to have 
a global ${\rm U(1)}_{\rm DM}$ and interacts with the pseudo scalar 
via Yukawa interaction while a $\mathbb{Z}_2$ symmetry 
is imposed on the other component -- the scalar. These ensure
the stability of both the dark matter components. Although the Lagrangian of the
present model is CP conserving, however the CP symmetry breaks spontaneously
when the pseudo scalar acquires a vacuum expectation value (VEV).
 The scalar
component of the 
dark matter in the present model also develops a VEV on spontaneous 
breaking of the $\mathbb{Z}_2$ symmetry. Thus the various interactions 
of the dark sector and the SM sector are progressed through 
the mixing of the SM like Higgs boson, the pseudo scalar
Higgs like boson and the singlet scalar boson. We show that the observed gamma 
ray excess from the Galactic Centre, self-interaction of dark matter from 
colliding clusters as well as the 3.55 keV X-ray line from Perseus, Andromeda 
etc. can be simultaneously explained in the present two component dark matter 
model. 
\end{abstract}
\newpage
\section{Introduction}  
\label{int}

The observational results from the satellite borne experiment
WMAP \cite{Hinshaw:2012aka} and more recently Planck \cite{Ade:2013zuv} have now
firmly established
the presence of dark matter (DM) in the Universe. Their results
reveal that more than 80\% matter content of the Universe
are in the form of mysterious unknown matter called the dark matter.
Until now, only the gravitational interactions of DM have
been manifested by most of its indirect evidences namely the flatness of
rotation curves of spiral galaxies \cite{Sofue:2000jx}, gravitational
lensing \cite{Bartelmann:1999yn}, phenomena of Bullet cluster
\cite{Clowe:2003tk} and other various colliding galaxy clusters etc.
However, the particle nature of DM still remains an enigma.
There are various ongoing dark matter direct detection
experiments such as LUX \cite{Akerib:2016vxi}, XENON-1T \cite{Aprile:2015uzo},
$\text{PandaX-II}$ \cite{Tan:2016zwf}
etc. which have been trying to investigate the particle nature 
as well as the interaction type (spin dependent or spin independent)
of DM with the visible sector by measuring the
recoil energy of the scattered detector nuclei.
However, the null results of these experiments have
severely constrained the DM-nucleon spin independent
scattering cross-section and thereby at present,
$\sigma_{\rm SI}>2.2\times 10^{-46}$ cm$^2$
has been excluded by the LUX experiment \cite{Akerib:2016vxi}
for the mass of a 50 GeV dark matter
particle at 90\% C.L. Like the spin independent
case, the present upper bound on DM-proton
spin dependent scattering cross-section is $\sigma_{\rm SD}
\sim 5\times 10^{-40}$ cm$^2$ \cite{Amole:2015pla,Amole:2016pye}
for a dark matter of mass $\sim 20$ to 60 GeV. 
The DM-nucleon scattering cross-sections are approaching
towards the regime of coherent neutrino-nucleon scattering
cross-section and within next few years $\sigma_{\rm SI}$
may hit the ``neutrino floor''. Therefore, it will be
difficult to discriminate the DM signal from that
of background neutrinos. However, if the DM is detected
in direct direction experiments then that will be
a ``smoking gun signature'' of the existence of beyond
Standard Model (BSM) scenario as the Standard Model of
particle physics does not have any viable cold dark matter
candidate. 

Depending upon the production mechanism at the
early Universe, the dark matter can be called
thermal or non-thermal. In the former case,
dark matter particles were in both thermal
as well as chemical equilibrium with other particles
in the thermal soup at a very early epoch. However,
the number density of DM became exponentially suppressed
(or Boltzmann suppressed) as the temperature of
the Universe drooped below the dark matter
mass ($T_{\rm Universe}\la M_{\rm DM}$)
which resulted in a reduced interaction rate
(interaction rate directly proportional to number density).
Decoupling of DM from the thermal bath occurred
at around a temperature $\sim \frac{M_{\rm DM}}{20}$
when DM interaction rate became subdominant
compared to the expansion rate of the Universe.
The corresponding temperature is known as
the freeze-out temperature of DM. After
decoupling DM became a {\it thermal relic}
with a {\it constant} density known as its relic density.  
Weakly Interacting Massive particle (WIMP)
\cite{Gondolo:1990dk,Srednicki:1988ce} is
the most favourite class for the thermal dark matter
scenario. Some of the most studied WIMPs in the existing
literature are neutralino \cite{Jungman:1995df}, scalar singlet
dark matter \cite{Silveira:1985rk}-\cite{Barger:2007im},
inert doublet dark matter \cite{Ma:2006km}-\cite{Banik:2014cfa}, singlet
fermionic dark matter \cite{Kim:2008pp}-\cite{Fairbairn:2013uta}, hidden sector
vector dark matter \cite{Hambye:2008bq}-\cite{DiChiara:2015bua} etc.  

On the other hand, in the non-thermal
scenario, the interaction strengths of DM particles
were so feeble that they never entered into thermal
equilibrium with the other particles in the cosmic
soup. As the Universe began to cool down,
these types of particles were started to produce mainly
from the decay of some heavy unstable
particles at the early epoch. However, in principle
they could also be produced from the annihilation
of particles in the thermal bath, but with a subdominant
rate compared to the production from decay of heavy particles.
In this situation DM relic density is generated from
a different mechanism known as the Freeze-in \cite{McDonald:2001vt,Hall:2009bx}
which is in a sense a opposite process to the usual
Freeze-out mechanism. This type of DM particles
are often called the Feebly Interacting Massive
Particle or FIMP. Sterile neutrino produced
from the decay of some heavy scalars \cite{Merle:2014xpa}-\cite{Shakya:2015xnx}
or gauge bosons \cite{Biswas:2016bfo} is a very good candidate
of FIMP. Moreover, various FIMP type DM candidate in different extensions of the
Standard Model have been studied in Refs.
\cite{Yaguna:2011qn}-\cite{Biswas:2016iyh}. 

Besides the direct detection searches for dark matter,
another promising detection method of DM is to detect
the annihilation or decay products of dark matter
trapped in the heavy dense region of celestial objects
namely core of the Sun, Galactic Centre (GC), dwarf galaxies etc. These
secondary particles which can revel the information
about the particle nature of DM are gamma ray, neutrinos,
charged cosmic rays including electrons, positrons
protons and antiprotons etc. This is known as the
indirect detection of dark matter. Study of Fermi-LAT data \cite{Atwood:2009ez}
by independent groups \cite{Goodenough:2009gk}-\cite{Calore:2014nla} have
observed an excess of gamma ray in the
energy range 1-3 GeV which can be interpreted as a result of dark matter
annihilation in the region of GC. Detailed study of the excess by Calore et. al.
\cite{Calore:2014nla} also have reported that the gamma ray excess in 1-3 GeV
energy range can
be explained by dark matter annihilation into $b \bar b$ with annihilation
cross-section ${\langle \sigma v\rangle}_{b \bar b}=1.76^{+0.28}_{-0.27} \times
10^{-26}~{\rm cm}^3{\rm s}^{-1}$ at GC having mass $49^{+6.4}_{-5.4}$ GeV.
Excess in GC gamma ray can also be explained from the point sources 
considerations \cite{Lee:2015fea} or millisecond pulsars \cite{Bartels:2015aea}
as well. Study of dwarf
spheoridals (dSphs) by Fermi-LAT and Dark Energy Survey (DES) provides bound on
DM annihilation cross-section with DM mass, is in agreement with the GC excess
results for DM obtained from \cite{Ackermann:2015zua,Drlica-Wagner:2015xua}.
Recent observations of 45 dwarf satellite galaxies by Fermi-LAT and DES
collaboration
\cite{Fermi-LAT:2016uux} also do not exclude the possibility of DM origin of GC
gamma ray excess. Different particle physics model for dark matter are explored
in order to explain this 1-3 GeV gamma ray excess at GC
\cite{Boucenna:2011hy}-\cite{Biswas:2016ewm}.
Apart from the GC excess gamma ray, their is also another observation of
unidentified 3.55 keV X-ray line from the study of 73 galaxy cluster by Bulbul
et.al. \cite{Bulbul:2014sua} and Boyarsky et. al \cite{Boyarsky:2014jta}
obtained from XMM Newton observatory. This unknown X-ray line can be explained
as DM signal and several dark matter model are invoked to explain this
phenomena \cite{Krall:2014dba}-\cite{Cheung:2014tha}. There are also attempts
claiming that this 3.55 keV line can have some astrophysical origin
\cite{Jeltema:2014qfa, Carlson:2014lla}. Hitomi collaboration
\cite{Aharonian:2016gzq} also suggest molecular interaction in nebula is 
responsible for this 3.55 keV signal which also requires
further test to be confirmed. Study of colliding galaxy clusters can also
provide valuable information for dark matter self interaction. An earlier
attempt to calibrate the dark matter self interaction have been made by
\cite{Harvey:2013tfa}. Recently an updated measurement for DM self interaction
by Harvey et. al. \cite{Harvey:2015hha} have measured DM self interaction from
the observations of 72 galaxy cluster collisions. From their observation of
spatial off set in collisions of galaxy cluster, DM self interaction is found to
be $\sigma/m<0.47$ cm$^2/$g with 95\% confidence limit (CL). DM self interaction
observation from
Abell 3827 cluster performed by \cite{Kahlhoefer:2015vua} also suggests that
$\sigma/m\sim1.5$ cm$^2/$g. A study of dark matter self interaction by Campbell
et. al. \cite{Campbell:2015fra} have reported that a light DM of mass 
lesser than $0.1$ GeV,
whose production is followed by freeze in mechanism can explain the self
interaction results from Abell 3827 by \cite{Kahlhoefer:2015vua}. 

Hence, above results clearly indicate that both the results for GC excess 
(requires a heavier DM candidate) and DM self interaction (prefers a light DM) 
can be explained simultaneously only with a multi component dark matter model. 
Therefore, in order to explain the Galactic Centre gamma ray excess
and DM self interaction bound 
from colliding galaxy cluster in a single framework
of particle dark matter scenario,
we propose a two component dark matter model where
the Standard Model is extended by adding one extra singlet
scalar and a fermion. An additional pseudo scalar is also
introduced to the SM. The dark fermion has an 
additional global U(1)$_{\rm DM}$ symmetry which  prevents 
its interaction with SM fermions. Although this dark fermion
can interact with the pseudo scalar through a fermion pseudo scalar
interaction involving $\gamma_5$ operator.  
The Lagrangian of the pseudo scalar is so chosen that there 
can be no explicit CP violation; the CP symmetry can only be 
spontaneously broken when the pseudo scalar acquires a nonzero 
VEV. We show that, in this model, the dark fermion
can play the role of a WIMP type dark matter candidate.  
The other component namely the singlet scalar (assumed to be lighter DM
candidate) in the present two component model has a $\mathbb{Z}_2$ symmetry 
imposed on it to prevent its direct interaction with the SM particles.
This light scalar field can be a viable FImP (denoted as FImP instead of FIMP 
for being less massive) type dark matter candidate by assuming it has 
sufficiently tiny interaction strength with other particles in the model. Study 
of thermal two component dark matter has been performed in literatures
\cite{Biswas:2013nn}-\cite{Bian:2014cja}. There are also works relating non
thermal multi component dark matter models explored to address the GC
gamma ray excess or dwarf galaxy excess along with 3.55 keV X-ray results
\cite{Biswas:2015sva,Biswas:2015bca}. However, our present work deals with a two
different types of DM candidates namely a WIMP (i.e., thermal DM) and a
non-thermal DM candidate FImP. In order to compute the relic abundance of this 
``WIMP-FImP'' system, we have solved a coupled Boltzmann equation
involving both the dark fermion and singlet scalar and their self as well as 
mutual interactions. Since we are considering a WIMP type dark fermion
which interacts with SM particle via a pseudo scalar mediator and FImP type 
singlet scalar, we show that our model can easily evade all
the existing stringent bounds on DM-nucleon spin independent scattering cross- 
section. We find that besides satisfying the relic density criterion
and other relevant experimental bounds, the annihilation of dark fermion
to $b\bar{b}$ (through pseudo scalar mediator) final state at the Galactic 
Centre can explain the Fermi-LAT observed gamma ray excess while the light 
scalar FImP DM can easily reproduce the DM self interaction required to explain 
the spatial off set in the collision of different galaxy clusters as obtained 
from \cite{Harvey:2015hha,Kahlhoefer:2015vua}. In addition, we show that within 
the existing framework of ``WIMP-FImP'' DM, the FImP dark matter component can 
also be able to explain the XMM Newton observed 3.55 keV X-ray anomaly from its
decay to two photon final states via its tiny mixing with SM like Higgs boson.



The paper is organised as follows. The two component ``WIMP-FImP'' dark matter
model is developed in Sect.~\ref{model}. The multi component dark matter
Boltzmann equation in the present model is addressed in 
Sect.~\ref{relicdensity}.
In Sect.~\ref{lhc} we provide the bounds from collider physics. Dark matter
self interaction and bounds from 3.55 keV X-ray is discussed in
Sect.~\ref{selfint}.
Phenomenology of the two component dark matter model is explored in 
Sect.~\ref{res} along with direct detection measurements. The results for
GC gamma ray excess and DM self interaction is presented in Sect.~\ref{GCself}.
Finally in Sect.~\ref{conc} the paper is summarised with concluding remarks. 
\section{Two Component Dark Matter Model}
\label{model}
The two component dark matter model having a fermionic
component as well as a scalar component, considered in this work,
is a renormalisable extension of the Standard Model (SM) by a
real scalar field $S$, a singlet Dirac fermion $\chi$ and a
pseudo scalar field $\Phi$. Therefore, in the present scenario
the dark sector is composed of a Dirac fermion $\chi$
and a real scalar. The Dirac fermion is a singlet under
the SM gauge group and it has a global U(1)$_{\rm DM}$ charge.
This prevents $\chi$ to couple with any Standard Model fermions
which ensures its stability. One the other hand, we impose
a discrete $\mathbb{Z}_2$ symmetry on the real scalar field
$S$ which forbids the appearance of any term in the Lagrangian containing
odd number of $S$ field . The discrete symmetry $\mathbb{Z}_2$
breaks spontaneously when $S$ gets a vacuum expectation value
(VEV). Also, we have assumed that the Lagrangian is CP 
invariant and the CP symmetry is subjected to a spontaneous
breaking when the pseudo scalar acquires a VEV. After
the breaking of all the imposed symmetries (e.g. ${\rm SU}(2)_{\rm L}
\times{\rm U}(1)_{\rm Y}$, $\mathbb{Z}_2$ and CP) of the
Lagrangian through the VEVs of the scalar fields, the real
real components of $H$, $\Phi$ and $S$ will mix among each other.
The lightest one with suitable mass and sufficiently low
values of mixing angles with other scalars can serve
as the FImP component of dark matter.   

The Lagrangian of the model thus can be written as 
\bea
{\cal L} &=& {\cal L}_{\rm SM} + {\cal L}_{\rm DM} + 
{\cal L}_{\Phi} + {\cal L}_{\rm int}\,\, ,
\label{eq1}
\eea
where the Lagrangian for the SM particles including the usual
kinetic term as well as the quadratic and quartic terms for
the Higgs doublet $H$, is represented by $\mathcal{L}_{\rm SM}$.
As mentioned above, the dark sector Lagrangian ${\cal L}_{\rm DM}$
has two parts namely the fermionic and the scalar, which are given by,
\bea 
{\cal L}_{\rm DM} &=& {\bar{\chi}}(i\gamma^\mu{\partial}_\mu - m)\chi + 
{\cal L}_S\,\,, 
\label{eq2}
\eea
with
\bea
\mathcal{L}_S = \frac{1}{2} (\partial_{\mu} S)(\partial^{\mu} S) -
\frac{{\mu}_s^2}{2} S^2 - \frac{\lambda_s}{4} S^4\,.
\label{eq2.1}
\eea
The Lagrangian ${\cal L}_\Phi$ for the pseudo scalar boson $\Phi$
is given by 
\bea
{\cal L}_\Phi &=& \frac {1} {2} (\partial_\mu \Phi )^2 - 
\frac {\mu_{\phi}^2} {2} \Phi^2 - \frac {\lambda_{\phi}}{4} \Phi^4\,\, .
\label{eq3}
\eea
Note that the above Lagrangian (Eq. \ref{eq3}) does not have any term 
in odd power of $\Phi$. This is to make ${\cal L}_{\Phi}$ 
CP-invariant. In the interaction term contains
the Yukawa type interaction between pseudo scalar $\Phi$ and
Dirac fermion $\chi$. In addition to that, it also contains
all possible mutual interaction terms among the scalar fields
$H$, $\Phi$ and $S$. The interaction Lagrangian is given as 
\bea
{\cal L}_{\rm int} &=& -\,i\,g\,\bar{\chi}\gamma_5\chi\,\Phi -
V^{\prime}(H,\Phi,S)\,,
\label{eq4}
\eea  
where scalars and pseudo scalar mutual interaction terms
are denoted by $V^{\prime}(H,\,S,\,\Phi)$. The expression of $V^{\prime}$ is
given as 
\bea
V^{\prime}(H,\,S,\,\Phi) ={\lambda_{H\Phi}} H^\dagger H\,\Phi^2
+ {\lambda_{HS}} H^\dagger H\,S^2 
+ \lambda_{\Phi S} \Phi^2\,S^2 \,\, .
\label{eq5}
\eea 
Note that as in Eq.~\ref{eq4} we have Yukawa term
involving $\gamma_5$ only hence the Lagrangian is CP
invariant and does not contain any explicit CP symmetry breaking term.
Moreover it is also assumed in the model that 
the pseudo scalar $\Phi$ acquires a non-zero VEV. As a consequence of 
this assumption, the CP of the Lagrangian is broken spontaneously. 

After the spontaneous symmetry breaking of SM gauge symmetry, Higgs acquires 
a VEV, $v_1$ ($\sim$ 246 GeV) and the fluctuating scalar field 
about this minima ($v_1$) is denoted as $h$. Denoting $v_2$ to be the
VEV of the pseudo scalar $\Phi$ and $v_3$, the VEV that the singlet scalar 
$S$ is assumed to acquire, we have
\be
H = \frac {1} {\sqrt {2}} \left ( \begin{array}{c} 0 \\ v_1 + h \end{array}
\right )\, ,\,\, \Phi = v_2 + \phi \,\, , \,\, S = v_3 +s \, .
\label{eq6}
\ee
It is to be noted that the global U(1)$_{\rm DM}$ symmetry is conserved
even after the spontaneous symmetry breaking. Let us consider
the scalar potential term $V$
\begin{eqnarray}
V &=& \mu_{H}^2\,H^\dagger H + \lambda_{H}\,(H^\dagger H)^2
+ \frac {\mu_{\phi}^2} {2} \Phi^2 + \frac {\lambda_{\phi}}{4} \Phi^4
+ \frac{{\mu}_s^2}{2} S^2 + \frac{\lambda_s}{4} S^4 \nonumber \\
&& + {\lambda_{H\Phi}} H^\dagger H\,\Phi^2
+ {\lambda_{HS}} H^\dagger H\,S^2 
+ \lambda_{\Phi S} \Phi^2\,S^2 \,.
\label{potential}
\end{eqnarray}

After symmetry breaking, the scalar potential Eq. (\ref{potential})
takes the following form 
\bea
V&=& \dfrac{\mu_H^2} {2} (v_1 + h)^2 + \dfrac{\lambda_H} {4} (v_1 + h)^4 + 
\dfrac{\mu_\Phi^2} {2} (v_2 + \phi)^2 + \nonumber \\
&& \dfrac{\lambda_\Phi} {4} (v_2 + \phi)^4 + 
\dfrac{\mu_S} {2} (v_3 + s)^2 + \dfrac{\lambda_S} {4} (v_3 + s)^4 + \nonumber\\
&& \dfrac{\lambda_{H\Phi}} {2} (v_1 + h)^2 (v_2 + \phi)^2 + 
\dfrac{\lambda_{HS}}{2} (v_1 + h)^2 (v_3 + s)^2 +
\lambda_{\Phi S}(v_2 + \phi)^2 (v_3 + s)^2 \,\, .
\label{eq7}
\eea 

Using the minimisation condition that 
\be
\left (\dfrac{\partial V}{\partial h}\right),\,
\left(\dfrac{\partial V}{\partial \phi}\right),\,
\left ( \frac {\partial V} {\partial s} \right)\Bigg \vert_
{h = 0,\,\phi=0,\,s=0} 
=0 \,\, ,
\ee
we obtain the three following conditions 
\bea
\mu_H^2 + \lambda_H v_1^2 + \lambda_{H\Phi} v_2^2 + \lambda_{HS} v_3^2 
&=& 0  \nonumber \\
\mu_\Phi^2 + \lambda_\Phi v_2^2 + \lambda_{H\Phi} v_1^2 + 
2\lambda_{\Phi S} v_3^2 
&=& 0 \nonumber \\
\mu_S^2 + \lambda_S v_3^2 + \lambda_{HS} v_1^2 + 
2\lambda_{\Phi S} v_2^2 &=& 0\, .
\eea 
The mass mixing matrix with respect to the
basis $h$-$\phi$-$s$ can now be constructed by
evaluating  
$\frac {\partial^2 V} {\partial h^2}$, 
$\frac {\partial^2 V} {\partial \phi^2}$, 
$\frac {\partial^2 V} {\partial s^2}$,
$\frac {\partial^2 V} {{\partial h}{\partial \phi}}$,
$\frac {\partial^2 V} {{\partial h}{\partial s}}$,
$\frac {\partial^2 V} {{\partial s}{\partial \phi}}$
at $h=\phi=s=0$ and is obtained as
\bea
\mathcal{M}^2_{\rm scalar} &=& 2\,\left(\begin{array}{ccc} 
\lambda_H\,v_1^2 & \lambda_{H\Phi}\,v_1\,v_2 & \lambda_{HS}\,v_1\,v_3 \\
\lambda_{H\Phi}\,v_1\,v_2 & \lambda_\Phi\,v_2^2 & 2\lambda_{\Phi 
S}\,v_2\,v_3 \\
\lambda_{HS}\,v_1\,v_3 & 2\lambda_{\Phi S}\,v_2\,v_3 & \lambda_S\,v_3^2
\end{array}
\right).
\label{mat}
\eea 
Diagonalising the symmetric mass matrix (Eq. \ref{mat})
by a unitary transformation we obtain three eigenvectors $h_1$, $h_2$ and
$h_3$ which represent three physical scalars. Each of the new eigenstate 
is a mixture of old basis states $h$, $\phi$ and $s$ depending on
the mixing angles $\theta_{12}$, $\theta_{23}$ and $\theta_{13}$ i.e.
\bea
\begin{pmatrix}
h_1\\
h_2\\
h_3
\end{pmatrix}=U(\theta_{12}, \theta_{13}, \theta_{23})
\begin{pmatrix}
h\\
\phi\\
s
\end{pmatrix} \,,
\label{pmns}
\eea 
where $U(\theta_{12}, \theta_{23}, \theta_{13})$ is the usual
PMNS matrix with mixing angles are
$\theta_{12},\,\theta_{23},\,\theta_{13}$ and
complex phase $\delta=0$. In this work, we
choose $h_1$ as the SM like Higgs boson which
has been discovered few years ago by the LHC experiments
\cite{Aad:2012tfa,Chatrchyan:2012ufa} at CERN.
Therefore, throughout this work we keep the mass ($m_{1}$)
of $h_1 \sim 125.5$ GeV\footnote{We assume mass of physical scalars
$h_j$ to be $m_j,~j=1-3$.}. One the other hand as mentioned
at the beginning of this Section, we consider $h_2$ is also heavy and the
lightest scalar $h_3$ to be a component of dark matter (FImP candidate). For
simplicity, Eq.
\ref{pmns}
can be rewritten as 
\bea
\begin{pmatrix}
h_1\\
h_2\\
h_3
\end{pmatrix}=
\left( \begin{array}{ccc} a_{11}~~a_{12}~~a_{13} \\ a_{21}~~a_{22}~~a_{23} \\
a_{31}~~a_{32}~~a_{33}
\end{array} \right)
\begin{pmatrix}
h\\
\phi\\
s
\end{pmatrix} \,,
\label{pmns1}
\eea 
where $a_{ij}$ are elements of PMNS matrix.

Further, in order to obtain a stable vacuum we have the following
bounds on the quartic couplings 
\bea
\lambda_H,\, \, \lambda_\Phi,\,\, \lambda_S &>& 0 \nonumber \\
\lambda_{H\Phi} + \sqrt {\lambda_H \lambda_\Phi} & > & 0 \nonumber \\
\lambda_{HS} + \sqrt {\lambda_H \lambda_S} & > & 0 \nonumber \\
2\lambda_{\Phi S} + \sqrt {\lambda_\Phi \lambda_S} & > & 0 
\eea
and 
\bea 
&&\sqrt{2(\lambda_{H\Phi} + \sqrt {\lambda_H \lambda_\Phi})
(\lambda_{HS} + \sqrt {\lambda_H \lambda_S}) 
(2\lambda_{\Phi S} + \sqrt {\lambda_\Phi \lambda_S})}\nonumber \\
&&+\sqrt {\lambda_H \lambda_\Phi \lambda_S}
+\lambda_{H\Phi} \sqrt {\lambda_S} + \lambda_{HS} \sqrt {\lambda_\Phi} + 
2\lambda_{\Phi S} \sqrt{\lambda_H}> 0 \,\, .
\label{vac}
\eea
In this model the fermionic dark matter (WIMP DM candidate) has an interaction
with the pseudo scalar $\Phi$ which should not be very large and be within the
perturbative limit. 
For this purpose we consider $g\le2\pi$ in our work.   
\section{Relic density}
\label{relicdensity}
\begin{figure}[!t]
\centering{
\includegraphics[height=12 cm, width=17 cm,angle=0]{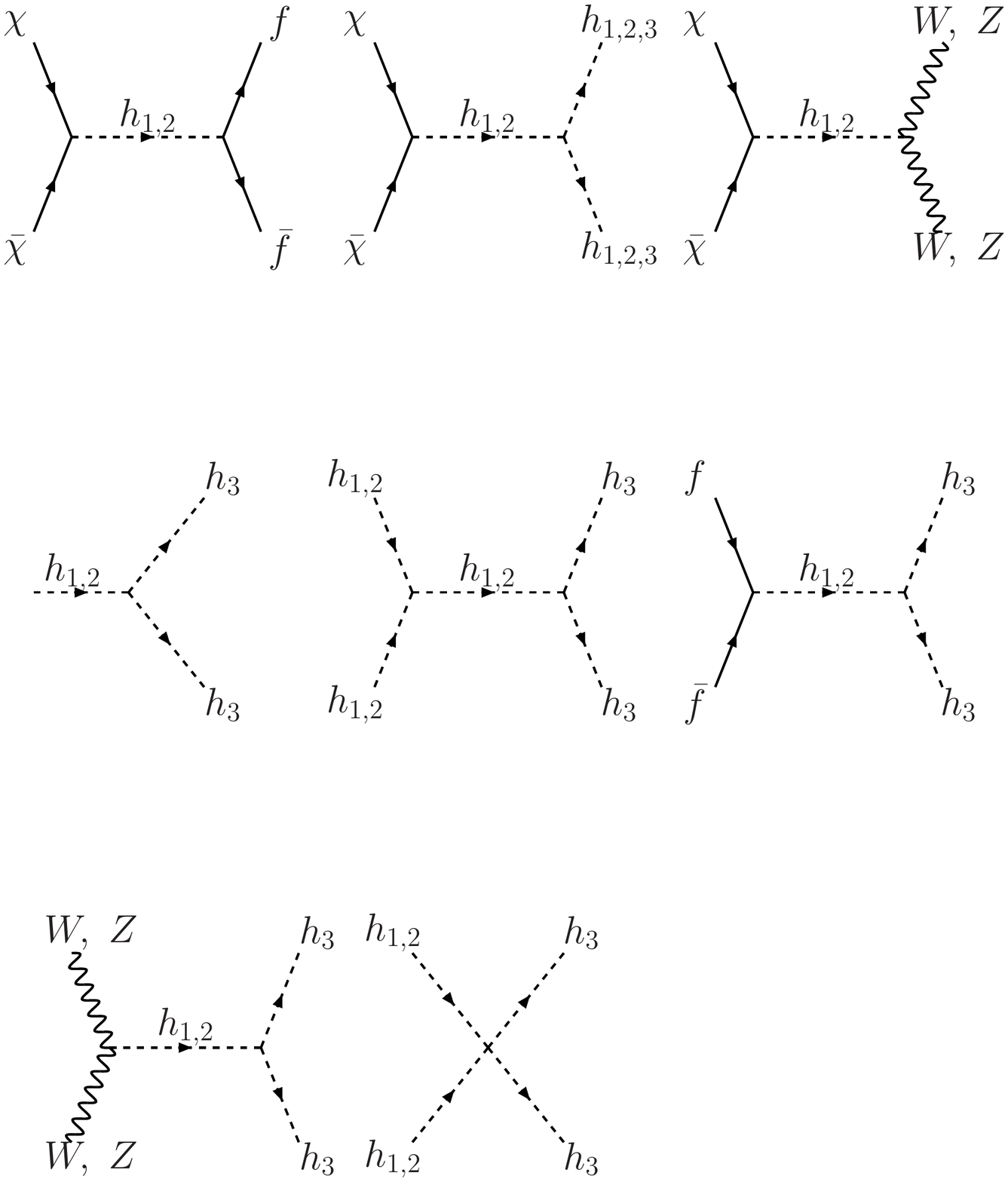}}
\caption{Feynman diagrams for the fermionic dark matter $\chi$ and scalar dark 
matter $h_3$}
\label{fig1}
\end{figure}

The relic density for the two component dark matter considered in the paper 
is obtained by solving the coupled Boltzmann equations for each of the dark 
matter components add then adding up the relic densities of each of the
components. 

The Boltzmann equation for the fermionic component $\chi$ in the present 
model is given by 
\bea
\frac {dY_{\chi}} {dz} &=& -\langle \sigma v \rangle_{\chi \chi \rightarrow x 
\bar{x}} 
\left ( Y_\chi^2 - (Y_\chi^{\rm eq})^2 \right ) + 
\langle \sigma v \rangle _{\chi\chi \rightarrow h_3h_3} 
\lt ( Y_\chi^2 - \frac {(Y_\chi^{\rm eq})^2} {(Y_{h_3}^{\rm eq})^2} Y_{h_3}^2 
\rt)
\,\, .
\label{bef}
\eea 
The fermionic dark matter in the present model follows usual freeze out
mechanism and becomes relic which behaves as a WIMP dark matter. However,
evolution of light dark matter $h_3$ is different. We assume that the mixing 
between the scalar $h_j, j=1-3$ are very small. Therefore the scalar $h_3$ is 
produced from the decay or annihilation heavier particles such as Higgs or 
gauge bosons which never reaches thermal equilibrium (therefore becomes 
non-thermal in nature) and its production 
saturates as the Universe expands and cools down. This is also referred as
freeze in production of particle \cite{McDonald:2001vt,Hall:2009bx} and the
light dark matter resembles a FImP like DM. Hence, initial abundance of $h_3$, 
$Y_{h_3} =0$ in the present model. Thus Eq. \ref{bef} takes the form  
\bea 
\frac {dY_{\chi}} {dz} &=& -\langle \sigma v \rangle_{\chi \chi \rightarrow x 
\bar{x}}
\left ( Y_\chi^2 - (Y_\chi^{\rm eq})^2 \right ) +
\langle \sigma v \rangle _{\chi\chi \rightarrow h_3h_3}
Y_\chi^2 
\,\, ,
\label{bef1}
\eea
where $x=f,~W,~Z,~h_1,~h_2$, denotes the final state particles produced due
to annihilation of dark matter candidate $\chi$.
The Boltzmann equation for the scalar component $h_3$ in the present 
framework is given by
\begin{eqnarray}
\frac{dY_{h_{{}_{{}_3}}}}{dz} &=& 
-\frac{2M_{pl}z}{1.66 m^2} 
\frac{\sqrt{g_{\star}(T)}}{g_{\rm s}(T)}
\Bigg(\sum_i\langle \Gamma_{h_{{}_{{}_i}} \rightarrow h_{{}_{{}_3}}
h_{{}_{{}_3}}} \rangle  
\left(Y_{h_{{}_{{}_3}}} - {Y^{eq}_{h_{{}_{{}_i}}}}\right)
\Bigg)\,
- \nonumber \\ && ~\frac{4 \pi^2}{45} 
\frac{M_{pl} m}{1.66}
\frac{\sqrt{g_{\star}(T)}}{z^2} \times
\nonumber \\ &&
\Bigg(\sum_{x = W, Z, f, h_1, h_2}
\langle {\sigma {\rm v}}_{x\bar{x}\rightarrow h_{{}_{{}_3}} h_{{}_{{}_3}}}
\rangle
\,\,{({Y}^2_{h_3}} -{Y^{eq}_{x}\,^2)} \,
+~\langle {\sigma {\rm v}}_{\chi_{{}_{{}}}\chi_{{}_{{}}}\rightarrow
h_{{}_{{}_3}} h_{{}_{{}_3}}} \rangle Y^2_{\chi}
\Bigg)\,\,.\nonumber \\
\label{bes}
\end{eqnarray}
With $Y_{h_3} = 0$, Eq. \ref{bes} takes the form
\begin{eqnarray}
\frac{dY_{h_{{}_{{}_3}}}}{dz} &=& 
-\frac{2M_{pl} z}{1.66 m^2} 
\frac{\sqrt{g_{\star}(T)}}{g_{\rm s}(T)}
\Bigg(\sum_i\langle \Gamma_{h_{{}_{{}_i}} \rightarrow h_{{}_{{}_3}}
h_{{}_{{}_3}}} \rangle  
\left(- {Y^{eq}_{h_{{}_{{}_i}}}}\right)
\Bigg)\,
- \nonumber \\ && ~\frac{4 \pi^2}{45} 
\frac{M_{pl} m}{1.66}
\frac{\sqrt{g_{\star}(T)}}{z^2} \times
\nonumber \\ &&
\Bigg(\sum_{x = W, Z, f, h_1, h_2}
\langle {\sigma {\rm v}}_{x\bar{x}\rightarrow h_{{}_{{}_3}} h_{{}_{{}_3}}}
\rangle
\,\,( -{Y^{eq}_{x}\,^2)} \,
+~\langle {\sigma {\rm v}}_{\chi_{{}_{{}}}\chi_{{}_{{}}}\rightarrow
h_{{}_{{}_3}} h_{{}_{{}_3}}} \rangle Y^2_{\chi}
\Bigg)\,\,.\nonumber \\
\label{bes1}
\end{eqnarray}
In Eqs. \ref{bef}-\ref{bes1}, $Y_x=\frac{n_x}{S}$ is the comoving number 
density of dark matter candidate $x=\chi,~h_3$ while $Y^{eq}_x$ is the 
equilibrium number density, $z=m/T$ where $T$ is the photon temperature and $S$
is the entropy of the Universe. $M_{pl}=1.22\times 10^{22}$ GeV in Eqs. 
~\ref{bes}-\ref{bes1} denotes the Planck mass and the term $g_{\star}$ is 
expressed as \cite{Gondolo:1990dk}
\begin{eqnarray}
\sqrt{{g_\star}(T)}&=&
\frac{g_S(T)}{\sqrt{g_{\rho}(T)}}
\left(1+\frac{1}{3}\frac{d~{\rm ln} g_S(T)}{d~{\rm ln} T}\right)
\label{gstar}
\end{eqnarray}
where $g_S$ and $g_{\rho}$ are the degrees of freedom corresponding to
entropy and energy density of Universe and written as \cite{Gondolo:1990dk}
\bea
S=g_S(T)\frac{2\pi^4}{45}T^3\, ,\hskip 15 pt
\rho=g_{\rho}(T)\frac{\pi^2}{30}T^4\,\,.
\label{density}
\eea
Thermal average of various annihilation cross-section ($\langle \sigma v 
\rangle$) and decay widths ($\langle \Gamma \rangle$) are given as
\bea
\langle \sigma {\rm v} \rangle_{aa\rightarrow 
bb}&=&\frac{1}{8m_a^4\,T\,K_2^2(m_a/T)}
\int_{4m_a^2}^\infty ds~\sigma_{aa\rightarrow bb}(s)~(s-4m_a^2)\,\sqrt{s}\,
K_1(\sqrt{s}/T)\nonumber \\
\langle \Gamma_{a\rightarrow bb} \rangle&=& \Gamma_{a\rightarrow 
bb}\frac{K_1(z)}{K_2(z)}\,\, .
\label{average}
\eea
In Eq. \ref{average} $K_1$ and $K_2$ are modified Bessel functions and $s$ 
represents the centre of momentum energy. Using Eq.~\ref{bef1},\ref{bes1}
and Eqs.~\ref{gstar}-\ref{average} we solve for the relic abundance of dark 
matter candidates given as
\bea
\Omega_{j}h^2= 2.755 \times 10^8 \left(\frac{m_j}{GeV}\right)Y_j(T_0),\hskip 10 
pt j=\chi,h_3
\label{relic}
\eea
where $T_0$ is the present photon temperature and $h$ is Hubble parameter
expressed in the unit of 100 km~s$^{-1}$~Mpc$^{-1}$. It is to be noted that
relic densities of these two dark matter components must satisfy the condition 
for total dark matter density obtained from Planck \cite{Ade:2013zuv} when added
up, i.e.,
\be
\Omega_{DM}h^2=\Omega_{\chi}h^2+\Omega_{h_3}h^2\,\, ,
\hskip 10 pt 0.1172\le \Omega_{DM}h^2 \le 0.1226\,\, .
\label{relic2}
\ee
Expressions of different annihilation cross-sections and decay processes 
along with the relevant couplings are given in Appendix A. Feynman
diagrams that contribute to the annihilations of $\chi$ along with the
production of scalar dark matter $h_3$ via decay and annihilation channels are
shown in Fig.~\ref{fig1}. It is to be noted that the diagram $\chi \chi
\rightarrow h_3 h_3$ will also contribute to the production of light scalar dark
matter. 

\section{Bounds from Collider Physics}
\label{lhc}
ATLAS and CMS have confirmed their observation of a Higgs like scalar with mass
$\sim$ 125.5 GeV \cite{Aad:2012tfa,Chatrchyan:2012ufa}. In the present model
described in Sect.~\ref{model}, we
introduced three scalar particles. As mentioned earlier we assume $h_1$ as the
Higgs like scalar and $h_2$ to be the non SM scalar ($85 {\rm GeV}\le m_2\le
110$ GeV) while $h_3$ is the light dark matter candidate. Since $h_1$ is the
Higgs like scalar with mass $\sim$ 125.5 GeV, we expect it to satisfy the 
collider
bounds on signal strength of SM scalar. We define signal strength as
\bea
R_1 =  \frac {\sigma (pp\rightarrow h_1)} {\sigma^{\rm SM}(pp\rightarrow h)}
\frac {{\rm Br} (h_1 \rightarrow xx)} {{\rm Br}^{\rm SM}(h \rightarrow xx)} 
\,\, .
\label{r1} 
\eea
In the above, $\sigma (pp\rightarrow h_1)$ defines the production cross-section
of $h_1$ due to gluon fusion while $\sigma^{\rm SM}(pp\rightarrow h)$ is the
same for SM Higgs. Similarly ${\rm Br} (h_1 \rightarrow xx)$ is defined as the
decay branching ratio of $h_1$ into any final particle whereas the same for SM
Higgs is ${\rm Br}^{\rm SM}(h \rightarrow xx)$. The Higgs like scalar must
satisfy the condition for SM Higgs signal strength signal $R_1\ge0.8$
\cite{ATLAS:2012xmd}. Branching ratio to any final state particle for $h_1$ is
given as ${\rm Br} (h_1
\rightarrow xx)=\frac{\Gamma(h_1\rightarrow xx)}{\Gamma_1}$ (here
$\Gamma(h_1\rightarrow xx)$ is decay width of $h_1$ into final state particles
and $\Gamma_1$ is the total decay width of $h_1$) and for SM Higgs with mass 
125.5 GeV it can be expressed as ${\rm Br}^{\rm SM}(h \rightarrow
xx)=\frac{\Gamma(h\rightarrow xx)}{\Gamma_{SM}}$, where $\Gamma_{SM}$ is total
decay width of Higgs. Hence, Eq.~\ref{r1} can be written as
\bea
R_1=a_{11}^4\frac{\Gamma_{SM}}{\Gamma_1}
\label{r_1}
\eea
where $\Gamma_1=a_{11}^2\Gamma_{SM} + \Gamma^{inv}_1$ is the total decay width 
and $\Gamma^{inv}_1$ is the invisible decay width of $h_1$ into dark matter 
particles given as
\bea
\Gamma^{inv}_1=\Gamma_{h_1\rightarrow \chi \bar{\chi}} + \Gamma_{h_1 \rightarrow
h_3 h_3}\,\, .
\label{inv1}
\eea
Similarly for $h_2$, the signal strength can be written as 
\bea
R_2=a_{21}^4\frac{\Gamma'_{SM}}{\Gamma_2}
\label{r_2}
\eea
with 
$\Gamma_2=a_{21}^2\Gamma'_{SM}+\Gamma^{inv}_2$ respectively where 
$\Gamma'_{SM}$ is the total decay width of non SM scalar of mass $m_2$ and 
$\Gamma^{inv}_2=\Gamma_{h_2\rightarrow \chi \bar{\chi}} + \Gamma_{h_2\rightarrow
h_3 h_3}$. The expression of invisible decay $\Gamma(h_i \rightarrow \chi 
\bar{\chi}),~i=1,2$ is
\bea
\Gamma_{h_1 \rightarrow \chi \bar{\chi}}=
\frac{m_1}{8\pi}g^2a_{21}^2
\left(1-\frac{4m_{\chi}^2}{m_1^2}\right)^{1/2}\,\, ,\nonumber \\
\Gamma_{h_2 \rightarrow \chi
\bar{\chi}}=\frac{m_2}{8\pi}g^2a_{22}^2
\left(1-\frac{4m_{\chi}^2}{m_2^2}\right)^{1/2}\,\, , 
\eea 
while the expression for $\Gamma_{h_j \rightarrow h_3 h_3},~j=1,2$ are given in
Appendix A. The invisible decay branching ratio for the SM like Higgs is 
$Br_{inv}^{1}=\frac{\Gamma_1^{inv}}{\Gamma_1}$. We assume the invisible decay
branching ratio to be small and impose the condition $Br_{inv}^{1}<0.2$
\cite{Belanger:2013kya}.
\section{Dark matter self interaction}
\label{selfint}
Study of dark matter self interaction have recently received attention and have
been explored in literatures
\cite{Harvey:2013tfa,Harvey:2015hha,Kahlhoefer:2015vua}. Dark matter, though
primarily thought to be collisionless in nature, is found to have self
interaction from the observation of colliding galaxy clusters. A study of 72
colliding clusters by Harvey et. al. \cite{Harvey:2015hha} claim that dark
matter self interaction cross-section $\sigma_{DM}/m<0.47$ cm$^2/$g with 95\%
CL. In the present model we proposed two dark matter
candidates $\chi$ (WIMP like fermion) and a light scalar dark matter $h_3$ 
(FImP). In this work
we will investigate whether any of these dark matter candidate can account for
the observed dark matter self interaction cross-section. Study of dark
matter self interaction by Campbell et. al. \cite{Campbell:2015fra} 
have reported that a light dark matter with mass below 0.1 GeV produced by
freeze in mechanism can provide the required amount of dark matter self
interaction cross-section (contact interaction) in order to explain the
observations of Abell 3827 \cite{Kahlhoefer:2015vua} with $\sigma_{DM}/m\sim1.5$
cm$^2/$g which is close to the bound obtained from \cite{Harvey:2015hha}.
Therefore in the present work, we investigate whether the
FImP dark matter $h_3$ (produced via freeze in mechanism 
as mentioned earlier in Sect.~\ref{relicdensity}) can account for the dark 
matter self interaction cross-section given by 
\cite{Harvey:2015hha,Kahlhoefer:2015vua}. The ratio
to self interaction cross-section with mass $m_3$ for the scalar dark matter
candidate in the present model is given as \cite{Campbell:2015fra}
\bea
\frac{\sigma_{h_3}}{m_3}=\frac{9\lambda_{3333}^2}{2\pi m_3^3} \,\, ,
\label{self}
\eea
where $\lambda_{3333}$ is the quartic coupling for $h_3$ given in Appendix A. In
the Eq.~\ref{self} we have considered contact interaction only and neglected 
the contributions from $s$-channel mediated
diagrams since those are suppressed due to small coupling with scalars $h_1$ and
$h_2$ and also due large mass terms in propagator. 
\subsection{3.55 keV X-ray emission and light dark matter candidate}
\label{xray}
Independent study of XMM Newton observatory data by Bulbul et. al.
\cite{Bulbul:2014sua} and Boyarsky et. al. \cite{Boyarsky:2014jta} have reported
a 3.55 keV X-ray emission line from 
extragalactic spectrum. Such an observation can not be explained by known 
astrophysical phenomena. Although the signal is not confirmed, if it remains 
to exist then such a signature can be explained by decay of heavy dark 
matter candidates \cite{Modak:2014vva} or annihilation of light dark matter
directly
into photon \cite{Biswas:2015sva,Babu:2014pxa}. The observations from Hitomi
collaboration
\cite{Aharonian:2016gzq} also 
suggests that the 3.55 keV X-ray line can be the caused by charge exchange 
phenomena in molecular nebula which requires more sensitive observation to be 
confirmed. Since in the present framework, we propose a light dark matter 
candidate $h_3$ to circumvent the self interaction property of dark matter, we 
further investigate whether it can also explain the 3.55 keV X-ray signal. For 
this purpose, we assume that mass of the light FImP dark matter candidate $h_3$
is $m_3\sim$ 7.1 keV which annihilate into pair of photons.

The expression for the decay of $h_3$ into 3.55 keV X-rays is given as
 \begin{equation}
\Gamma_{h_3 \rightarrow \gamma \gamma} =
\left(\frac{\alpha_{\rm em}}{4 \pi}\right)^2 |F|^2 \,\,
a_{31}^2 \, \frac{ G_F m^3_{3}}{8 \sqrt{2} \pi} \,\, , 
\label{x}
\end{equation}
where $G_F$ is the Fermi constant and $\alpha_{\rm em}\sim\frac{1}{137}$ is the 
fine structure constant. The loop factor $F$ in Eq.~\ref{x} is
\begin{equation}
F = F_W(\beta_W) + \sum_fN_c Q_f^2 F_f(\beta_f)
\end{equation}
where
\begin{eqnarray*}
\beta_W &=& \frac{4 m_W^2}{m^2_{3}},
~~\beta_f = \frac{4 m_f^2}{m^2_{3}},\nonumber \\
F_W(\beta) &=& 2+3\beta+3 \beta(2-\beta) f(\beta),\nonumber\\
F_f(\beta) &=& -2\beta[1+(1-\beta)f(\beta)],\nonumber\\
f(\beta) &=& \arcsin^2[\beta^{-1/2}]~.
\end{eqnarray*}
$N_c$ in the loop factor is the colour quantum number while $Q_f$ denotes the 
charge of the fermion. It is to be noted that the decay width of $h_3$ must be 
in the range $2.5\times 10^{-29}~{\rm s^{-1}}\le f_{h_3}\Gamma_{h_3 
\rightarrow \gamma \gamma} \le 2.5\times 10^{-28}~{\rm s^{-1}}$ in order to 
produce the required extragalactic X-ray flux obtained from Andromeda, Perseus 
etc. Since in the present model we have two dark matter components, the decay 
width of $h_3$ must be multiplied by a factor 
$f_{h_3}=\frac{\Omega_{h_3}}{\Omega_{DM}}$, is the fractional contribution to 
dark matter relic density by $h_3$ component. Hence, in this work we will 
also test the viability of the light scalar dark matter candidate to explain 
the possible X-ray emission signal reported by 
\cite{Bulbul:2014sua,Boyarsky:2014jta} along with DM self interaction results.
\section{Calculations and Results}
\label{res}

\begin{figure}[h!]
\centering
\subfigure[]{
\includegraphics[height=5.5 cm, width=5.5 cm,angle=0]{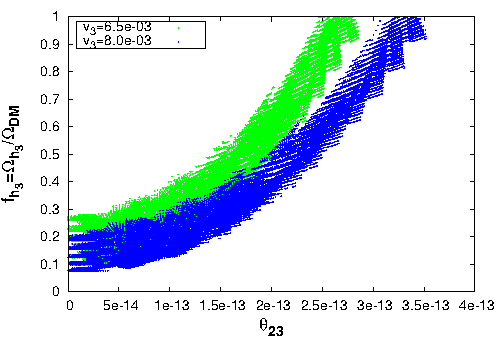}}
\subfigure []{
\includegraphics[height=5.5 cm, width=5.5 cm,angle=0]{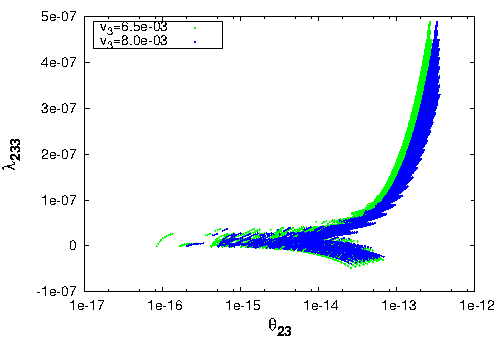}}
\subfigure []{
\includegraphics[height=5.5 cm, width=5.5 cm,angle=0]{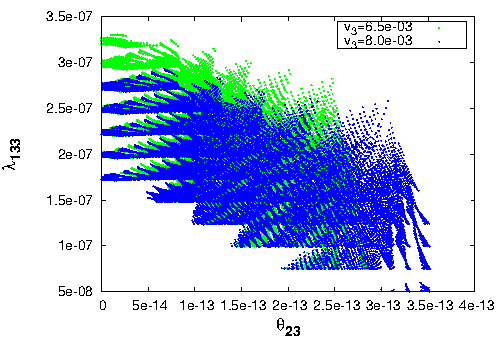}}
\caption{The left panel (Fig.~\ref{fig2}a) shows the changes in $f_{h_3}$
with mixing angle $\theta_{23}$. Fig.~\ref{fig2}b-c depicts the allowed values
of the couplings $\lambda_{233}$ and $\lambda_{133}$ plotted
against $\theta_{23}$.}
\label{fig2}
\end{figure}

\begin{table}
\begin{center}
\resizebox{\textwidth}{!}{
\begin{tabular}{|c|c|c|c|c|c|c|c|c|c|}
\hline
 $m_1$ & $m_2$ & $m_3$ & $\lambda_{12}$ & $\lambda_{13}$ & $\lambda_{23}$ & 
$R_1$ & $Br_{inv}^{1}$ & $f_{h_3}\Gamma_{h_3 \rightarrow \gamma \gamma}$&$g$ 
\\
GeV&GeV&GeV& & & & & & 10$^{-29}$~s$^{-1}$&\\
\hline
$\sim$125.5&85-110&$\sim$7.1$\times$10$^{-6}$&10$^{-4}$-0.1&10$^{-10}$-10$^{-8}
$& 10$^{-11}$-10$^{-9}$&0.8-1.0&0-0.2&2.5-25&0.01-5.0\\
\hline
\end{tabular}
}
\end{center}
\caption{Constraints and chosen region of model parameters space for the two 
component dark matter model.}
\label{tab1}
\end{table}

In this section we test the viability of the present two component dark matter
model scanning over a range of model parameter space. In Table~\ref{tab1}, we
tabulate the range of model parameter space and relevant constraints used in 
this work. Note that the coupling parameters $\lambda_{ij};~i,j=1-3,~(i\neq j)$
are in agreement with the vacuum stability conditions mentioned earlier in 
Eq.~\ref{vac} (Sect.~\ref{model}) and also satisfy perturbative unitarity 
condition. As we have mentioned earlier, $h_1$ is SM like scalar and $h_2$ is 
non SM scalar, we take $v_1=246$ GeV and $v_2=500$ GeV in the model. We further 
assume two choices of $v_3=$ 6.5 MeV and 8.0 MeV. This choice is consistent with
the previous studies of light scalar dark matter of mass$\sim$7.1 keV with 
bound 2.0 MeV $\le v_3 \le$ 10.0 MeV \cite{Biswas:2015sva,Babu:2014pxa}. We have
also imposed
the conditions on signal strength and invisible decay branching ratio of 
SM like scalar $h_1$ obtained from ATLAS and CMS at LHC ($R_1\ge 0.8$ and
$Br_{inv}^{1}\le0.2$). Using the range of model parameter space tabulated in
Table~\ref{tab1} we solve the three scalar mass mixing matrix in order 
to find out the elements of PMNS matrix $a_{ij};~i.j=1-3$ (and mixing angle). 
These matrix 
elements are then used to calculate various couplings mentioned in Appendix A
which are necessary in order to calculate the decay widths and annihilation
cross-sections of scalar dark matter candidate $h_3$. The coupling $g$ ($\le
2\pi$, bound from perturbative limit) between the pseudo scalar and the 
fermionic
dark matter is also varied within the range mentioned in Table~\ref{tab1} to
compute the annihilation cross-sections for fermionic dark matter. These decay
widths and annihilation cross-sections of both dark matter candidates are then
used to solve for the coupled Boltzmann Eqs.~\ref{bef1},\ref{bes1} and
calculate the relic densities for each dark matter candidate satisfying the
condition for total dark matter relic density Eq.~\ref{relic2}. In
Fig.~\ref{fig2} we show valid range of model parameter space obtained using
Table~\ref{tab1} and solving the coupled Boltzmann equations satisfying the
condition $\Omega_{\chi}h^2+\Omega_{h_3}h^2=\Omega_{DM}h^2$ as given by Planck
satellite experiment. In Fig.~\ref{fig2}a we plot the variation of allowed
mixing angles $\theta_{23}$ with the fractional relic density $f_{h_3}$ of the
scalar dark matter in the present framework \footnote{Mixing angles
$\theta_{ij};~i,j=1-3,i\neq j$ are expressed in radian.}. Plotted blue and
green shaded
regions depicted in all the three figures of Fig.~\ref{fig2} corresponds to the
choice of $v_3=6.5\times10^{-3}$ GeV and $8.0\times10^{-3}$ GeV. The
observation of Fig.~\ref{fig2}a (in $\theta_{23}-f_{h_3}$ plane) shows that the
relic density
contribution of the scalar dark matter component increases with the increase in
$\theta_{23}$. It is to be noted that the maximum allowed range of $\theta_{23}$
depends on the choice of $v_3$
and we have found that for $v_3=6.5\times 10^{-3}$ GeV $\theta_{23}^{max}\sim
2.8\times 10^{-13}$ while the same obtained with  $v_3=8.0\times 10^{-3}$ GeV is
$\theta_{23}^{max}\sim 3.5\times 10^{-13}$. This variation of $\theta_{23}$
with $f_{h_3}$ shown in Fig.~\ref{fig2}a is a direct consequence of the fact
that increase in $\theta_{23}$ also increases the value of $\lambda_{233}$ which
is depicted in Fig.~\ref{fig2}b. In Fig.~\ref{fig2}b the variation of
$\theta_{23}$ is plotted against $\lambda_{233}$. It is easily seen from
Fig.~\ref{fig2}b that when $\theta_{23}$ is small $\sim 10^{-16}-10^{-14}$, the
value of $\lambda_{233}$ is very small. However as $\theta_{23}$ increase
further, there is a sharp increase in the value of $|\lambda_{233}|$. As a
result the contribution from the decay channel $h_2 \rightarrow h_3 h_3$
enhances which then also raises the relic density contribution of scalar $h_3$.
From Fig.~\ref{fig2}b we notice that maximum allowed range of $\lambda_{233}$
is $\sim 5 \times 10^{-7}$ for both the cases of $v_3$ considered in the work.
Finally in Fig.~\ref{fig2}c $\theta_{23}$ is plotted against $\lambda_{133}$ for
the both the values of $v_3$ mentioned above. From Fig.~\ref{fig2}c we notice
that $\lambda_{133}$ decreases steadily with enhancement in $\theta_{23}$
indicating an suppression in the contribution from $h_1$ (with $m_1\sim 125.5$
GeV) decay into pair of $h_3$. The allowed range of $\lambda_{133}$ for both
the values of $v_3$ lie within the range $0.5\times 10^{-8}-3.5\times 10^{-7}$. 
In the present work mass of $h_2$
is varied in the range $85-110$ GeV (i.e., $m_2<m_1$) and decay width is
inversely proportional to the mass of decaying particle (see Appendix A for
expression). This indicates that the contribution of the non SM scalar to the
freeze in production of FImP dark matter $h_3$ is significant compared to the
same obtained from SM like scalar when coupling $\lambda_{233}$ is not small
(i.e., $ |\lambda_{233}|\sim\lambda_{133}$).

\begin{figure}[h!]
\centering
\subfigure[]{
\includegraphics[height=5.5 cm, width=5.5 cm,angle=0]{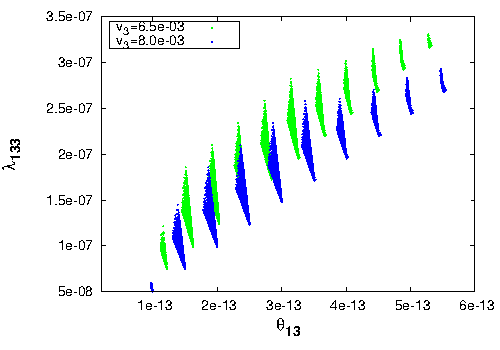}}
\subfigure []{
\includegraphics[height=5.5 cm, width=5.5 cm,angle=0]{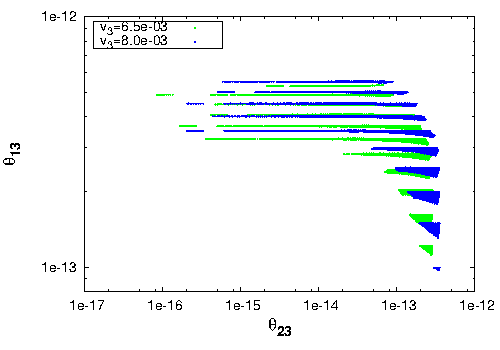}}
\caption{The available model parameter space in $\theta_{13}-\lambda_{133}$
plane is shown in the left panel (Fig.~\ref{fig3}a) while in the right panel
(Fig.~\ref{fig3}b) the same region is depicted when $\theta_{23}$ is varied
against $\theta_{13}$.}
\label{fig3}
\end{figure}

Fig.~\ref{fig3}a depicts the allowed range of $\theta_{13}$ plotted against
$\lambda_{133}$ for both the values of $v_3$ considered in earlier plots of
Fig.~\ref{fig2}. We also use the similar color scheme to indicate the values of
$v_3$ satisfying the same conditions applied in order to plot
Fig.~\ref{fig2}. From Fig.~\ref{fig3}a it can be easily observed that
$\theta_{13}$ in the present model varies within the range $\sim 1.0-6.0\times
10^{-13}$ for both the chosen values of $v_3=6.5\times 10^{-3}$ GeV and
$v_3=8.0\times 10^{-3}$ GeV respectively. It can also noticed from the plots in
Fig.~\ref{fig3}a that $\lambda_{133}$ is proportional to the value of
$\theta_{13}$. This reveals that the decay width $h_1 \rightarrow h_3 h_3$
increases with increase in $\theta_{13}$ which can enhance the freeze in pair
production of $h_3$ via $h_1$. In Fig.~\ref{fig3}b we show the allowed model
parameter space in $\theta_{23}-\theta_{13}$ plane for the same set of $v_3$
values and constrains used in earlier plots as well. Study of Fig.~\ref{fig3}b
reveals that 
for smaller values of $\theta_{23}\sim 10^{-16}-10^{-14}$, $\theta_{13}$
maintains a value in range $\sim 3\times 10^{-13}-6\times 10^{-13}$ indicating
that contribution in the relic density is mostly contributed from the decay of
$h_1$ into two $h_3$ scalars. However, as $\theta_{23}$ increase the
contribution of $h_2$ increases (due to increase in $\lambda_{233}$) which
reduces the value of $\theta_{13}$ (as well as $\lambda_{133}$) in
order to maintain the contribution to total DM relic density by $h_3$ and to
avoid overabundance of dark matter (when we add up the contribution on DM relic
density obtained from the fermionic dark matter component $\chi$, i.e., 
$f_{h_3}+f_{\chi}=1$).
It is to be mentioned that the mixing angles $\theta_{12}$ varies within the 
range $0.003\le \theta_{12}\le 0.183$ for the allowed model parameter space 
obtained using both set of $v_3$ considered. Note that all the 
plots in Fig.~\ref{fig2} and 
Fig.~\ref{fig3} are in agreement with the constraints on decay width of 7.1 keV 
scalar $h_3$ into X-ray, $2.5\times 10^{-29}$s$^{-1}\le f_{h_3}\Gamma_{h_3 
\rightarrow \gamma \gamma}\le~2.5\times10^{-28}$s$^{-1}$. We have also found
that the signal strength of $h_2$, i.e., $R_2$ in the present formalism is 
very small to be observed at the LHC experiments due to smallness of mixing 
between SM like scalar $h_1$ with $h_2$. 
\begin{figure}[h!]
\centering
\subfigure[]{
\includegraphics[height=5.5 cm, width=5.5 cm,angle=0]{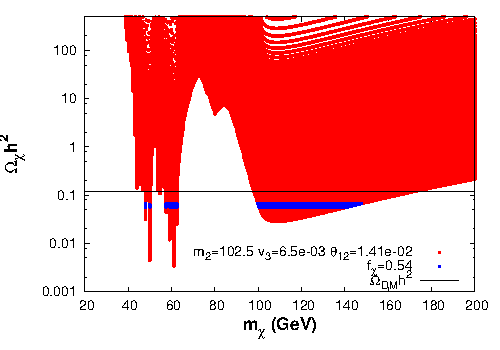}}
\subfigure []{
\includegraphics[height=5.5 cm, width=5.5 cm,angle=0]{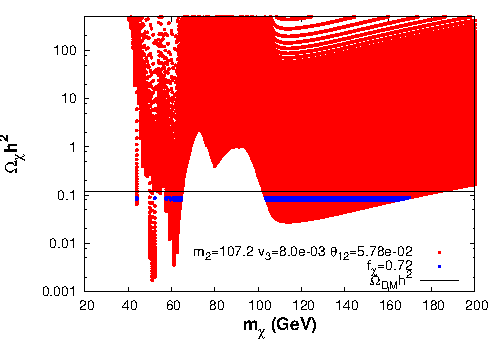}}
\subfigure []{
\includegraphics[height=5.5 cm, width=5.5 cm,angle=0]{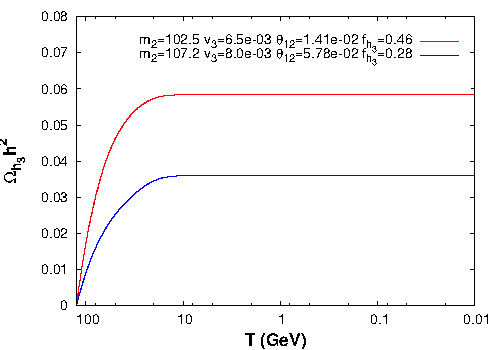}}
\caption{Plots in Fig.~\ref{fig4}a-b shows the $m_{\chi}-\Omega_{\chi}h^2$
parameter space for the set of parameters in table~\ref{t2} for the fermionic
DM. The variation of $\Omega_{h_3}h^2$ (for the scalar DM $h_3$) with
temperature $T$ for the same set of parameter is shown in Fig.~\ref{fig4}c.}
\label{fig4}
\end{figure}

\begin{table}
\begin{center}
\begin{tabular}{|c|c|c|c|c|c|c|}
\hline
Set&$m_1$ & $m_2$ & $m_3$ & $v_{3}$ & $\theta_{12}$&$g$ 
\\
&GeV&GeV&GeV&GeV&& \\
\hline
1&125.4&102.5&7.12$\times10^{-6}$&6.5$\times10^{-3}$&1.41$\times10^{-2}$&
0.01-5.0\\
\hline
2&125.5&107.2&7.15$\times10^{-6}$&8.0$\times10^{-3}$&5.78$\times10^{-2}$&
0.01-5.0\\
\hline
\end{tabular}
\end{center}
\caption{Chosen parameter set for the plots in Fig.~\ref{fig4}a-c.}
\label{t2}
\end{table}

So far, in this work, we have only discussed about the available parameters for
the two 
component dark matter model involving a fermion $\chi$ and a light scalar $h_3$
of mass $\sim 7.1$ keV in agreement with Planck dark matter relic density
satisfying the condition $\Omega_{\chi}h^2+\Omega_{h_3}h^2=\Omega_{DM}h^2$
(Fig.~\ref{fig2}-\ref{fig3}). In Fig.~\ref{fig4}a-b we show the
$m_{\chi}-\Omega_{\chi}h^2$ plots while in Fig.~\ref{fig4}c the variation of
dark matter density $\Omega_{h_3}h^2$ for light dark matter candidate $h_3$
($m_3\sim 7.1$ keV) is plotted against the temperature $T$ of Universe.
Instead of scanning over the full range of parameter space obtained from
Fig.~\ref{fig2} and Fig.~\ref{fig3} (for two values of $v_3$), we consider two
valid set of parameters for the purpose of demonstration tabulated in
Table~\ref{t2}. Therefore, the parameter sets in Table~\ref{t2} is within the
range of scan performed using the Table~\ref{tab1} and also respects all other
necessary conditions (such as vacuum stability, decay width of $h_3$,
constrains from LHC etc.). Fermionic dark matter candidate can annihilate
through $s$-channel annihilation mediated by scalars $h_1$ and $h_2$ (see
Fig.~\ref{fig1}). The mixing between the SM like scalar $h_1$ and non SM scalar
$h_2$ given by $\theta_{12}$, is necessary to calculate the parameters
$a_{ij},~i,j=1,2$ and different annihilations of the fermionic dark matter.
Since in the present work the range of coupling
$\lambda_{12}$ is larger compared to other couplings $\lambda_{23}$ and
$\lambda_{13}$, the parameters $a_{ij},~i,j=1,2$ will dominantly be determined
by $\theta_{12}$. This is also justified by the plots in Fig.~\ref{fig3}b where 
$\theta_{23}$ is varied with $\theta_{13}$ showing these mixing angles are 
very small. Therefore, we have chosen two values 
of $\theta_{12}$ for two set of $v_3$ values given in Table~\ref{t2}. 
Note that we have also considered the same set of $v_3$ values of light scalar 
$S$ in our model along with $v_1=246$ GeV and $v_2=500$ GeV taken earlier in 
order to find out the valid range parameter space obtained in 
Figs.~\ref{fig2}-\ref{fig3}. Shown $m_{\chi}-\Omega_{\chi}h^2$ plot in
Fig.~\ref{fig4}a corresponds to the set of parameters with 
$v_3=6.5\times10^{-3}$
GeV and the same with other set of parameters (when $v_3=8.0\times10^{-3}$ GeV)
is depicted in Fig.~\ref{fig4}b. The red regions in both the Figs.~\ref{fig4}a-b
is obtained by varying the coupling $g$ within the range $0.01\le g\le5.0$ and
also varying the fermionic dark matter mass $m_{\chi}$ from 20 GeV to 200 GeV.
From both the Figs.~\ref{fig4}a-b it can be observed that a very small region of
parameter space (for these chosen sets in Table~\ref{t2}) lies below the total
dark matter density bound given by Planck \cite{Ade:2013zuv}
(black horizontal line shown in both the plots Fig.~\ref{fig4}a-b). We have
found that
relic density of fermionic dark matter becomes less abundant with respect to
total dark matter relic density near the resonances of SM like Higgs ($h_1$)
and non SM scalar $h_2$ when its mass $m_{\chi}\sim m_{i}/2,~i=1,2$.
Apart from that, there is also a region of parameter space with mass $\sim
100-180$ GeV (for $v_3=6.5\times10^{-3}$ GeV) and $\sim100-190$ GeV (when
$v_3=8.0\times10^{-3}$ GeV) where the condition
$\Omega_{\chi}h^2<\Omega_{DM}h^2$ is satisfied. In this region the heavy
fermionic dark matter annihilates into scalar $h_1$ and $h_2$. Thus the dark
matter annihilation cross-section get enhanced which reduces the relic density
$\Omega_
{\chi}h^2$ of fermionic dark matter candidate. Shaded blue horizontal
regions shown in the plot Fig.~\ref{fig4}a (Fig.~\ref{fig4}b) are fractional
contributions to the total DM relic density from fermionic dark matter candidate
$\chi$ with $f_{\chi}=0.54$ ($f_{\chi}=0.72$) where 
$f_{\chi}=\frac{\Omega_{\chi}}{\Omega_{DM}}$. In Fig.~\ref{fig4}c we show the
evolution of relic density $\Omega_{h_3}h^2$ of the light scalar dark matter
$h_3$ as a function of temperature $T$ of the Universe with the same set of
parameters given in
Table~\ref{t2}. The plot shown in red (blue) depicted in Fig.~\ref{fig4}a
(Fig.~\ref{fig4}b) corresponds to the parameter set with $v_3=6.5\times10^{-3}$
GeV ($v_3=8.0\times10^{-3}$ GeV). Moreover, we have also satisfied the condition
$f_{\chi}+f_{h_3}=1$ in the plots of Fig.~\ref{fig4}c (in order to produce the 
total DM relic abundance obtained from Planck results \cite{Ade:2013zuv}) such
that the fractional
contribution of $h_3$ for each set of parameter in Table~\ref{t2} is
$f_{h_3}=1-f_{\chi}$, i.e., $f_{h_3}=0.46~(0.28)$ for the red (blue) plot
depicted in Fig.~\ref{fig4}c. It appears from the plots in Fig.~\ref{fig4}c
that the relic density of light scalar dark matter is very small (as initial
abundance $Y_{h_3}=0$), increases gradually with decreasing temperature and
finally saturates near $T\sim$ 10 GeV. The saturation of the relic density
indicates that the production of $h_3$ ceases as the Universe expands and cools
down due to rapid decrease in the number density of decaying or annihilating
particles. Therefore from Fig.~\ref{fig4}a-c it can be concluded that the
present model of two component dark matter with a WIMP (heavy fermion $\chi$)
and a FImP (light scalar $h_3$) can successfully provide the observed dark
matter relic density predicted by Planck satellite data.
\subsection{Direct detection of dark matter}
\label{dd}
In this section we will investigate whether the 
allowed model parameter space is compatible with the results from direct 
detection of dark matter obtained from dark matter direct detection experiments.
Direct detection experiments search for the evidences of dark matter-nucleon 
scattering and provides bounds on dark matter-nucleon scattering cross-section. 
Dark matter candidates in the present model can undergo collision with detector
nucleus and the recoil energy due to the scattering is calibrated. Since no 
such collision event have been observed yet by different dark matter direct 
detection experiments, these experiments provide an exclusion limit on dark 
matter-nucleon 
scattering cross-section. The most stringent bound on DM-nucleon 
spin independent (SI) cross-section 
is given by LUX \cite{Akerib:2016vxi}, XENON-1T \cite{Aprile:2015uzo} and
PandaX-II \cite{Tan:2016zwf}. In the
present model both the dark matter components (WIMP and FImP)
$\chi$ and $h_3$ can suffer spin independent (SI) elastic scattering with the 
detector nucleus. The fermionic dark matter $\chi$ in the present work can 
interact 
through pseudo scalar interaction via $t$-channel processes mediated by both 
$h_1$ and $h_2$. The expression of spin independent scattering cross-section
for the fermionic dark matter $\chi$ is
\bea
\sigma_{SI}^{\chi}=\frac{g^2}{\pi}m_r^2
\left(\frac{a_{11}a_{12}}{m_1^2}+\frac{a_{22}a_{21}}{m_2^2}\right)^2
\lambda_p^2~v^2
\label{fermdd}
\eea
where $\lambda_p$ is given as \cite{LopezHonorez:2012kv}
\be
\lambda_p=\frac{m_p}{v_1}\left[\sum_q f_q+ \frac{2}{9}\left(1-\sum_q
f_q\right)\right] \simeq1.3\times10^{-3} \,\, .
\label{lp}
\ee
and $m_r=\frac{m_{\chi}m_p}{m_{\chi}+m_p}$ denotes the reduced mass for the 
scattering. It is to be noted that due to the pseudo scalar interaction 
scattering cross-section on Eq.~\ref{fermdd} is velocity suppressed and 
hence multiplied by a factor $v^2~$ with $v\sim 10^{-3}$ being the velocity of 
dark matter particle. We have found that this velocity suppressed scattering 
cross-section is way below the latest limit on DM-nucleon scattering given by 
Direct detection experiments \cite{Akerib:2016vxi}-\cite{Tan:2016zwf} DM direct
search
experiment. This finding is also in agreement with the results obtained in a
different work by Ghorbani \cite{Ghorbani:2014qpa}. Moreover, since we 
have two dark matter components in the model, the effective scattering 
cross-section for the fermionic dark matter (i.e., WIMP candidate) will be
rescaled by a factor 
proportional to the fractional number density 
$r_{\chi}=\frac{n_{\chi}}{n_{\chi}+n_{h_3}}$ ($n_x$ denotes the number 
density), i.e., $\sigma_{SI}^{'\chi}=r_{\chi}\sigma_{SI}^{\chi}$ (for further
details see \cite{Biswas:2014hoa,Biswas:2015sva}). The number 
density of both the dark matter components $\chi$ and $h_3$ can be obtained 
from the expression of individual relic density given in Eq.~\ref{relic}. In the 
present framework the fermionic dark matter candidate $\chi$ is $\sim 10^{6}$ 
times heavier than the scalar $h_3$ dark matter. For example if we consider 
that the contribution to the total relic density from $h_3$ is smaller with 
respect to that of fermion $\chi$ having value $\Omega_{h_3}h^2\sim 
0.1\Omega_{\chi}h^2$, the number density of $h_3$ is $10^{6}$ times larger than 
that of $n_{\chi}$. This 
indicates that the rescaling factor $r_{\chi}\sim 10^{-6}$ and $r_{h_3}\sim 1$.
Therefore the effective spin independent scattering cross-section 
$\sigma_{SI}^{'\chi}$ for fermionic dark matter candidate is further suppressed 
by the rescaling factor $r_{\chi}<<1$ making it much smaller than the 
most sensitive dark matter direct detection limits obtained from experiments 
like LUX, PandaX-II. 
Similarly, for the scalar FImP dark matter candidate the effective spin
independent 
direct detection cross-section is given 
as $\sigma_{SI}^{'h_3}= r_{h_3}\sigma_{SI}^{h_3}$ where
\be
\sigma_{SI}^{h_3}=\frac{m_r^{'2}}{4\pi}\frac{f^2}{v_1^2}\frac{m_p^2}{m_3^2}
\left(\frac{\lambda_{133}a_{11}}{m_1^2}+\frac{\lambda_{233}a_{21}}{m_2^2}
\right)^2
\,\, ,
\label{h3dd}
\ee
where $m'_r=\frac{m_{3}m_p}{m_{3}+m_p}$ and $f\sim$0.3 \cite{hall}. Since 
$m_3<<m_p$, $m'_r\sim m_3$ and Eq. \ref{h3dd} can be rewritten as
\be
\sigma_{SI}^{h_3}=\frac{1}{4\pi}\frac{f^2}{v_1^2}m_p^2
\left(\frac{\lambda_{133}a_{11}}{m_1^2}+\frac{\lambda_{233}a_{21}}{m_2^2}
\right)^2
\,\, .
\label{h3dd1}
\ee
Since $h_3$ in the present model has very small interaction with the SM bath
particles and never reaches equilibrium after once produced, the couplings 
$\lambda_{133}$ and $\lambda_{233}$ are very small ($\sim10^{-7}$, as seen from 
Fig.~\ref{fig2}b-c). We have found that though the number density of $h_3$ is 
high $r_{h_3}\sim 1$ (as it is light), effective scattering cross-section 
$\sigma_{SI}^{'h_3}\sim\sigma_{SI}^{h_3}$ is also very small to be observed by
any dark matter direct search experiments and remains far below the most
stringent limit given by LUX \cite{Akerib:2016vxi}, XENON-1T
\cite{Aprile:2015uzo} and  PandaX-II \cite{Tan:2016zwf} due to smallness of
couplings 
$\lambda_{j33},~j=1,2$. Therefore, in the present scenario of two component dark
matter model (with a WIMP and a FImP), we do not expect any bound on model
parameter space from direct detection experimental constraints. 


\section{Galactic Centre gamma ray excess and dark matter self interaction}
\label{GCself}
An excess of gamma ray in the energy range 1-3 GeV have been obtained from
analysis of Fermi-LAT data \cite{Atwood:2009ez} in the region of Galactic
Centre. 
Such an excess can be interpreted as a result of dark matter annihilation in 
the GC region. Dark matter particles can be trapped due the immense 
gravitational pull of GC and 
also other astrophysical sites like dwarf galaxies, Sun etc. These sites are 
rich with particle dark matter which then undergo pair annihilation. Different 
particle physics models for dark matter are explored in order to provide a 
suitable explanation 
to this excess in gamma ray at GC as we have mentioned earlier in
Sect.~\ref{int}. An analysis of this 1-3 GeV GC excess
gamma ray by Calore, Cholis and Weniger (CCW) \cite{Calore:2014nla} using
various galactic diffusion excess models suggests that Fermi-LAT data can be 
explained by dark matter annihilation at GC. Indeed, the $\gamma$-ray excess
can be very well fitted with a dark matter of mass $49^{+6.4}_{-5.4}$ GeV
which annihilates into pair of $b \bar{b}$ particles \footnote{Produced pair of 
fermions undergo hadronisation processes to finally annihilate into pair of 
photons via pion decay or bremsstrahlung.} with annihilation 
cross-section ${\langle \sigma v\rangle}_{b \bar b}=1.76^{+0.28}_{-0.27} 
\times 10^{-26}~{\rm cm}^3{\rm s}^{-1}$. In this section we will 
investigate whether the WIMP like fermionic dark matter candidate $\chi$ can
account for the observed GC gamma ray excess results. In addition, self
interaction study of the light scalar dark matter (FImP DM, mentioned earlier in
Sect.~\ref{selfint}) will also 
be addressed in this section. Before we explore the dark matter interpretation 
of GC gamma ray excess, a discussion is in order. The study of gamma ray 
signatures from dwarf galaxies by Fermi-LAT and DES
\cite{Ackermann:2015zua,Drlica-Wagner:2015xua} also 
provide limits on dark matter annihilation cross-section into various 
annihilation modes. The limits on dark matter annihilation cross-section into 
$b \bar{b}$ is consistent with the GC gamma ray excess analysis by CCW. 
However, apart from dark matter annihilation, the gamma ray excess at GC in the 
range 1-3 GeV can also be explain by various non DM phenomena such as 
contribution from point sources near GC \cite{Lee:2015fea} or millisecond
pulsars \cite{Bartels:2015aea}. Study by Clark et. al. \cite{Clark:2016mbb} 
also rule out the idea that the point like sources are dark matter
substructures. However, in a recent work Fermi-LAT and DES collaboration have
performed an analysis of $\gamma$-ray data with 45 confirmed dwarf spheroidals
(dSphs) \cite{Fermi-LAT:2016uux}. The analysis of gamma ray emission data from
these dSphs by Fermi-LAT and DES provides bound on dark matter annihilation
cross-section into different final channel particles ($b \bar{b}$ and $\tau
\bar{\tau}$). Although their analysis \cite{Fermi-LAT:2016uux} of the data
do not show any significant excess at these sites (dSphs), the limits
obtained on DM annihilation cross-section in their analysis do not exclude the
possibility of DM interpretation of GC gamma ray excess either. Therefore in the
present work, we will consider dark matter as the source to the gamma ray 
excess at Galactic Centre observed by Fermi-LAT and test the viability of our 
model.

\begin{table}
\begin{center}
\begin{tabular}{|c|c|c|c|c|c|c|c|c|c|c|c|}
\hline
BP1& $m_1$ & $m_2$ & $m_{\chi}$ &$v_3$&$g$ & $R_1$ & $Br_{inv}^{1}$ & 
$f_{\chi}$&$f_{\chi}^2{\langle \sigma v \rangle}_{b \bar{b}}$& $r_{\chi}$&
$\sigma_{SI}^{'\chi}$\\
 &GeV&GeV&GeV&$10^{-3}$& & & & &10$^{-26}$& & pb\\
 &   &   &   &   GeV   & & & & &cm$^3$s$^{-1}$& &\\
\hline
1&125.9&104.3&50.0&6.5&0.07&0.89&0.079&0.89&1.66&1.19e-06&2.39e-26\\
\hline
2&125.8&106.8&47.5&8.0&0.05&0.94&0.038&0.91&1.54&1.36e-06&1.15e-26\\
\hline
\end{tabular}
\end{center}
\caption{Benchmark points for calculation of GC gamma ray excess plotted in
Fig.~\ref{fig5} with fermionic dark matter $\chi$.}
\label{t3}
\end{table}

\label{gc}
\begin{figure}[h!]
\centering
{\includegraphics[height=6.0 cm, width=7.5 cm,angle=0]{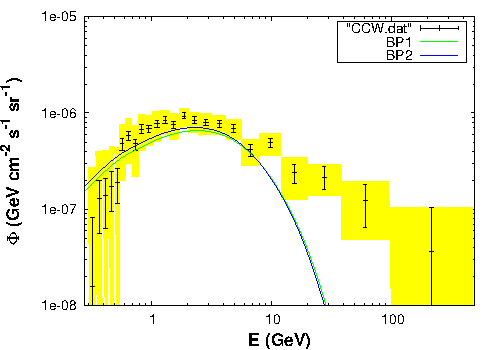}}
\caption{Comparison of the Fermi-LAT excess results from CCW 
\cite{Calore:2014nla} with the gamma ray flux obtained from benchmark points 
tabulated in Table~\ref{t3}.}
\label{fig5}
\end{figure}
\begin{table}
\begin{center}
\begin{tabular}{|c|c|c|c|c|c|c|c|c|c|}
\hline
BP1& $m_1$ & $m_2$ & $m_3$ & $v_3$ &$f_{h_3}$ & $f_{h_3}\Gamma_{h_3 \rightarrow 
\gamma \gamma}$ & $r_{h_3}$ & $\frac{\sigma_{h_3}}{m_3}$& $\sigma_{SI}^{h_3}$\\
 &GeV&GeV&keV&$10^{-3}$& &10$^{-29}$ & &cm$^2/$g & pb\\
 &   &   &   &  GeV    & & s$^{-1}$  & &         &   \\
\hline
1&125.9&104.3&7.12&6.5&0.11&3.36&$\sim1$&0.313&7.08e-24 \\
\hline
2&125.8&106.8&7.15&8.0&0.09&5.06&$\sim1$&0.137&7.15e-24 \\
\hline
\end{tabular}
\end{center}
\caption{Calculations of different observables for the scalar dark matter 
candidate for the same set of benchmark points given in Table~\ref{t3}.}
\label{t4}
\end{table}

The expression for the differential gamma ray flux obtained a region of 
Galactic Centre for the fermionic dark matter candidate $\chi$ is 
\bea
\frac{\rm d^2\Phi}{\rm dE d\Omega}=\frac{{\langle \sigma {\rm 
v}\rangle}_f}{8\pi m_{\chi}^2}J
\frac{dN^{f}_{\gamma}}{dE_{\gamma}} \,\, ,
\label{gamflux}
\eea 
performed over a solid angle $d\Omega$ for certain region of interest (ROI).
From Eq.~\ref{gamflux}, it can be observed that the differential $\gamma$-ray
flux depends on the thermal averaged annihilation cross-section ${\langle 
\sigma v\rangle}_f$ of dark matter into final state particles 
(fermions) and $\frac{dN^{f}_{\gamma}}{dE_{\gamma}}$, is the photon energy 
spectrum produced due to per annihilation into fermions. 
In the above Eq.~\ref{gamflux}, the factor $J$, the astrophysical factor
depending on the dark matter density $\rho$, is expressed as
\bea
J=\int_{\rm los}\rho^2(r(r',\theta))dr\,\,  ,
\label{jfact}
\eea 
is the line of sight integral where 
$r'=\sqrt{r_{\odot}^2+r^2-2r_{\odot}r\cos\theta}$ with $r$ being the distance 
from the region of annihilation (GC) to Earth and $r_{\odot}=8.5$ kpc. The 
angle between line of sight and line from GC is denoted by $\theta$. In this 
work, we assume the dark matter distribution is spherically symmetric which 
follows Navarro-Frenk-White (NFW) \cite{Navarro:1996gj} profile given as
\bea
\rho(r)=\rho_s\frac{(r/r_s)^{-\gamma}}{(1+r/r_s)^{3-\gamma}}\, .
\label{halo}
\eea 
In the expression of NFW halo profile $r_s=20$ kpc and $\rho_s$ is a typical 
scale density such that it produce the local dark matter density 
$\rho_{\odot}=0.4$ GeV cm$^{-3}$ at a distance $r_{\odot}$. The differential 
gamma ray flux is calculated using the ROI used in the work  by CCW
\cite{Calore:2014nla} 
($|l|\le20^0$ and $2^0\le|b|\le20^0$) for $\gamma=1.2$. The photon spectrum 
$\frac{dN^{f}_{\gamma}}{dE_{\gamma}}$ from the annihilation of dark matter is
obtained from Cirelli \cite{Cirelli:2010xx}. In order to calculate the
differential gamma ray 
flux obtained for the fermionic dark matter using Eqs.~\ref{gamflux}-\ref{halo}
and the specified ROI by CCW, we consider two benchmark points from the 
available model parameter space discussed earlier in Sect.~\ref{res}. Therefore 
the benchmark points are in agreement with all the limits and constrains such 
as vacuum stability, LHC bounds, limits on decay width of light scalar, dark 
matter relic density etc. The benchmark points used to calculate gamma ray 
flux in this work is tabulated 
in Table~\ref{t3}. It is to be noted that since the dark matter candidate is 
fermion, one may think that the annihilation cross-section will be velocity 
suppressed. However, in the present model, the fermion dark matter has a 
pseudo scalar type interaction which removes the velocity dependence of dark 
matter annihilation cross-section \cite{Ghorbani:2014qpa}. In Fig.~\ref{fig5},
we compare the 
GC gamma ray flux produced using benchmark points BP1 and BP2 tabulated in 
Table~\ref{t3} with the results from CCW \cite{Calore:2014nla} for GC gamma ray
excess. 
It is to be noted that the annihilation cross-section for the fermionic dark 
matter $\chi$ into $b 
\bar{b}$, i.e., ${\langle \sigma v\rangle}_{b \bar b}$ will be multiplied by 
$f_{\chi}^2$ (since annihilation requires two dark matter
candidates)\footnote{This can be understood as the modified line of sight
integral $J_{eff}=f_{\chi}^2J$ as well depending on DM density.}. Hence in
order to produce the required flux for excess GC gamma ray, the contribution to 
the relic density by the fermionic candidate $f_{\chi}$ should be large. In 
Fig.~\ref{fig5}, the gamma ray flux obtained from BP1 (BP2) is plotted in 
green (blue)
along with the data obtained from CCW \cite{Calore:2014nla}. From
Fig.~\ref{fig5}, it can be 
observed that the fermionic dark matter component $\chi$ (WIMP) in our model
can account for the observed GC gamma ray excess results obtained by analysis
of Fermi-LAT data. Moreover, from the benchmark points it can also be seen that
the spin independent direct detection cross-section for the fermionic dark 
matter candidate calculated using Eqs.~\ref{fermdd},\ref{lp} is very small and 
remains below the limits from most stringent constraints on DM-nucleon
cross-section given by LUX \cite{Akerib:2016vxi}, XENON-1T \cite{Aprile:2015uzo}
etc. 

As we have 
mentioned earlier, we now investigate whether the light scalar dark matter $h_3$
can satisfy the condition for dark matter self interaction with the same set of
benchmark points. The relevant results for the scalar dark matter candidate 
$h_3$ for BP1 and BP2 are tabulated in Table.~\ref{t4}. From 
Table.~\ref{t4}, it can be easily seen that for both the benchmark points, 
the light scalar dark matter can provide a self interaction cross-section
consistent with the observed limits $\sigma/m\le 0.47$ cm$^2/$g  obtained from 
the study by Harvey et. al. \cite{Harvey:2015hha}\footnote{Although the 
contribution of scalar dark matter to the DM relic density is small, due to 
its small mass compared to the fermion candidate, the number density is huge.
This indicates that the self interaction process will mostly be attributed 
from the collisions of $h_3$ and effective self interaction 
$\frac{\sigma'_{h_3}}{m_3}=r^2_{\chi}\frac{\sigma_{h_3}}{m_3}\sim\frac{\sigma_{
h_3}}{m_3}$ since $r_{\chi}\sim1$.}. The self interaction for the light scalar 
DM candidate is calculated using Eq.~\ref{self}. 

It can also be seen from Table~\ref{t4} that the FImP like scalar DM can also
explain the 3.55 keV X-ray emission
as observed by XMM Newton observatory if confirmed later as well. Calculation 
of DM-nucleon scattering cross-section for the scalar dark matter (using 
Eq.~\ref{h3dd1}) also indicates that direct detection of the candidate is not 
possible at present having a small $\sigma_{SI}^{'h_3} $ compared to the upper 
limit obtained LUX and other DM direct search experiments. Hence, at present,
both the dark matter candidates ($\chi$ and $h_3$) are beyond reach of ongoing 
direct DM search experiments with spin independent scattering 
cross-section lying far below the existing limits obtained from these 
experiments. This justifies our previous comments on the scattering 
cross-section for the dark matter particles with detector nucleon discussed
in Sect.~\ref{dd}.
\section{Summary and conclusion}
\label{conc}
In this work we have explored the viability of a two component dark matter
model with a fermionic dark matter that evolve thermally behaving like a WIMP
and a non-thermal feebly interacting light singlet scalar dark matter which is
produced via freeze in mechanism (FImP). The
fermionic dark matter candidate $\chi$ interacts with the SM sector through a
pseudo scalar particle $\Phi$ as the pseudo scalar acquires a non zero VEV and
thus CP symmetry of the Lagrangian is broken spontaneously. Similarly the
$\mathbb{Z}_2$ symmetry of the singlet scalar is also broken spontaneously when
$S$ is given a tiny non-zero VEV resulting three physical scalars. However, the
global $U(1)_{\rm DM}$ symmetry of the fermionic dark matter remains intact to
provide us stable dark WIMP like DM candidate. On the other hand the light
scalar $h_3$ having a very small interaction with SM sector also serves as a
FImP dark matter candidate produced via freeze in mechanism. The $SU(2)_{\rm
L}\times
U(1)_{\rm Y}$ symmetry of SM Higgs field is also broken spontaneously which
provide mass to the SM particles. Hence, in the present model we have three
scalars which mix with each other. We identify one of the
physical scalar $h_1$ to be SM like, $h_2$ as non SM Higgs and $h_3$ is the
light scalar dark matter. We constrain the model parameter space by vacuum
stability, unitarity, bounds from LHC results on SM scalar etc. to solve for the
coupled Boltzmann equation in the present framework such that sum of relic
densities of these dark matter candidates satisfy the observed DM relic density
by Planck.
We test for the viability of fermionic dark matter candidate in order to
explain the GC gamma ray results obtained from the analysis of Fermi-LAT data
\cite{Atwood:2009ez} by CCW \cite{Calore:2014nla}. We show that excess of GC 
gamma ray in the energy range 1-3 GeV can be obtained from the annihilation of 
fermionic dark matter that produce the required amount of annihilation 
cross-section $\langle \sigma v\rangle_{b \bar{b}}$ having mass $\sim$ 50 GeV. 
There is also a valid region for the fermionic dark matter candidate $\chi$ 
with mass ranging from 100-190 GeV. In addition, we investigate whether the 
light scalar dark matter candidate can account for dark matter self 
interaction. We found that the light scalar dark matter $h_3$ considered in the 
model can provide the desired dark matter self interaction 
cross-section in order to explain the results from galaxy cluster collisions 
\cite{Harvey:2015hha,Kahlhoefer:2015vua}. Moreover, we also test for viability 
of this light dark matter candidate to explain the possible 3.55 keV X-ray 
signal obtained from the study of extragalactic X-ray emission reported by 
Bulbul et. al \cite{Bulbul:2014sua}. Our study reveals that a
light dark matter $m_3\sim 7.1$ keV in the present model can serve as a viable 
candidate that produce the required flux (in agreement with the condition for 
decay width $h_3 \rightarrow \gamma \gamma$) if confirmed by the observations 
of extragalactic X-ray search experiments and also consistent with the 
dark matter self interaction results.
Both the dark matter candidates in the
present ``WIMP-FImP'' framework are insensitive to direct detection experimental
bounds and spin independent direct detection cross-section is far below the
upper limit given by LUX DM direct search results. While this work is being
completed, we came to know about a new work \cite{Karwin:2016tsw} on analysis of
Fermi-LAT GC gamma ray excess for pseudo scalar interaction of dark matter using
a different ROI ($15^0\times15^0$) about GC with interstellar emission models
(IEMs) and point sources. A detailed study of the results presented in
\cite{Karwin:2016tsw} is beyond scope of this work and we wish to test these 
results for pseudo scalar interactions in our model in a future work.
\vskip 5mm
{\bf Acknowledgments} : Authors would like to thank P. Roy for his useful
suggestions and valuable discussions.

\vskip 3mm
\noindent {\bf Appendix A}
\vskip 2mm 
\begin{itemize}
\item Annihilation cross-section of fermion dark matter candidate $\chi$
\bea
\sigma {v}_{\chi \chi \rightarrow f \bar f} &=& N_c
\frac{g^2}{32\pi}
s\frac{m_f^2}{v_1^2}\left(1-\frac{4m_f^2}{s}\right)^{3/2}F(s,m_1,m_2)\,\, ,
\nonumber 
\label{chiff}
\eea
{\small
\bea
\sigma {v}_{\chi \chi \rightarrow W^+W^-} =
\frac{g^2}{64\pi}
\left(1-\frac{4m_W^2}{s}\right)^{1/2} \left(\frac{m_W^2}{v_1}\right)^2
\left(2+\frac{(s-2m_W^2)^2}{4m_W^4}\right)
F(s,m_1,m_2)\,\, ,\nonumber 
\label{chiww}
\eea}
{\small
\bea 
\sigma {v}_{\chi \chi \rightarrow ZZ} =\frac{g^2}{128\pi}
\left(1-\frac{4m_Z^2}{s}\right)^{1/2} \left(\frac{m_Z^2}{v_1}\right)^2
\left(2+\frac{(s-2m_Z^2)^2}{4m_Z^4}\right)
F(s,m_1,m_2)\,\, .\nonumber 
\label{chizz}
\eea}
{\small
\bea
F(s,m_1,m_2) &=& \left [\frac{a_{12}^2a_{11}^2}{(s-m_1^2)^2+m_1^2\Gamma_1^2} +
\frac{a_{21}^2a_{22}^2}{(s-m_2^2)^2+m_2^2\Gamma_2^2}
\right . \nonumber \\&& \left .
+a_{12}a_{11}a_{22}a_{21}\frac{2(s-m_1^2)(s-m_2^2)+2m_1m_2\Gamma_1\Gamma_2}
{[(s-m_1^2)^2+m_1^2\Gamma_1^2][(s-m_2^2)^2+m_2^2\Gamma_2^2]} \right]\,\,
.\nonumber
\eea}
{\small
\bea
\sigma {v}_{\chi \chi \rightarrow h_1h_1} &=& \frac{g^2}{32\pi}
\left(1-\frac{4m_1^2}{s}\right)^{1/2}
\left [\frac{a_{12}^2\lambda_{111}^2}{(s-m_1^2)^2+m_1^2\Gamma_1^2} 
\right . \nonumber \\&& \left .
+\frac{a_{22}^2\lambda_{211}^2}{(s-m_2^2)^2+m_2^2\Gamma_2^2}  
+\frac{2a_{12}a_{22}\lambda_{111}\lambda_{211}((s-m_1^2)
(s-m_2^2)+m_1m_2\Gamma_1\Gamma_2)}
{[(s-m_1^2)^2+m_1^2\Gamma_1^2][(s-m_2^2)^2+m_2^2\Gamma_2^2]} \right ]\,\, ,
\nonumber
\label{h1h1}
\eea}
{\small
\bea
\sigma {v}_{\chi \chi \rightarrow h_2h_2} &=& \frac{g^2}{32\pi}
\left(1-\frac{4m_2^2}{s}\right)^{1/2}
\left [\frac{a_{12}^2\lambda_{122}^2}{(s-m_1^2)^2+m_1^2\Gamma_1^2} 
\right . \nonumber \\&& \left .
+\frac{a_{22}^2\lambda_{222}^2}{(s-m_2^2)^2+m_2^2\Gamma_2^2}  
+\frac{2a_{12}a_{22}\lambda_{122}\lambda_{222}((s-m_1^2)
(s-m_2^2)+m_1m_2\Gamma_1\Gamma_2)}
{[(s-m_1^2)^2+m_1^2\Gamma_1^2][(s-m_2^2)^2+m_2^2\Gamma_2^2]} \right ]\,\, ,
\nonumber
\label{h2h2}
\eea}
{\small
\bea
\sigma {v}_{\chi \chi \rightarrow h_3h_3} &=& \frac{g^2}{32\pi}
\left(1-\frac{4m_3^2}{s}\right)^{1/2}
\left [\frac{a_{12}^2\lambda_{133}^2}{(s-m_1^2)^2+m_1^2\Gamma_1^2} 
\right . \nonumber \\&& \left .
+\frac{a_{22}^2\lambda_{233}^2}{(s-m_2^2)^2+m_2^2\Gamma_2^2}  
+\frac{2a_{12}a_{22}\lambda_{133}\lambda_{233}((s-m_1^2)
(s-m_2^2)+m_1m_2\Gamma_1\Gamma_2)}
{[(s-m_1^2)^2+m_1^2\Gamma_1^2][(s-m_2^2)^2+m_2^2\Gamma_2^2]} \right ]\,\, .
\nonumber
\label{h3h3}
\eea}
\item Decay and annihilation terms for scalar dark matter candidate $h_3$
\begin{eqnarray*}
\Gamma_{h_{{}_{{}_j}}\rightarrow h_{{}_{{}_3}}h_{{}_{{}_3}}} &=&
\frac{\lambda^2_{j33}}{8 \pi m_j}
\sqrt{1-\frac{4m^2_3}{m^2_j}}\,\, ,j=1,2\,\, ,
~~~~~~~~~\nonumber
\end{eqnarray*}
{
\bea
\sigma_{f \bar{f}\rightarrow h_3 h_3} =
N_c\frac{1}{16\pi s}\sqrt{(s-{4m_3^2})(s-4m_f^2)}
\left(\frac{m_f}{v_1}\right)^2
F'(s,m_1,m_2)\,\, ,\nonumber 
\label{ff33}
\eea}
{
\bea
\sigma_{W^+W^-\rightarrow h_3 h_3} =
\frac{1}{18\pi s}\sqrt{\frac{s-{4m_3^2}}
{s-4m_W^2}} \left(\frac{m_W^2}{v_1}\right)^2
\left(2+\frac{(s-2m_W^2)^2}{4m_W^4}\right)
F'(s,m_1,m_2)\,\, ,\nonumber 
\label{ww33}
\eea}
{
\bea 
\sigma_{ZZ\rightarrow h_3 h_3} =\frac{1}{18\pi s}
\sqrt{\frac{s-{4m_3^2}}{s-4m_Z^2}} \left(\frac{m_Z^2}{v_1}\right)^2
\left(2+\frac{(s-2m_Z^2)^2}{4m_Z^4}\right)
F'(s,m_1,m_2)\,\, .\nonumber 
\label{zz33}
\eea}
{
\bea
F'(s,m_1,m_2) &=& \left
[\frac{a_{11}^2\lambda_{133}^2}{(s-m_1^2)^2+m_1^2\Gamma_1^2} +
\frac{a_{21}^2\lambda_{233}^2}{(s-m_2^2)^2+m_2^2\Gamma_2^2}
\right . \nonumber \\&& \left .
+a_{11}\lambda_{133}a_{21}\lambda_{233}\frac{
2(s-m_1^2)(s-m_2^2)+2m_1m_2\Gamma_1\Gamma_2 }
{[(s-m_1^2)^2+m_1^2\Gamma_1^2][(s-m_2^2)^2+m_2^2\Gamma_2^2]} \right]\,\,
.\nonumber
\eea}
{
\bea 
\sigma_{h_1h_1\rightarrow h_3h_3} =\frac{1}{2\pi s}
\sqrt{\frac{s-{4m_3^2}}{s-4m_1^2}} 
\left(\lambda_{1133}+3\frac{\lambda_{111}\lambda_{133}}{(s-m_1^2)}
+\frac{\lambda_{211}\lambda_{233}}{(s-m_2^2)}\right)^2\,\, .\nonumber 
\label{1133}
\eea}
{
\bea 
\sigma_{h_2h_2\rightarrow h_3h_3} =\frac{1}{2\pi s}
\sqrt{\frac{s-{4m_3^2}}{s-4m_2^2}} 
\left(\lambda_{2233}+3\frac{\lambda_{222}\lambda_{233}}{(s-m_2^2)}
+\frac{\lambda_{122}\lambda_{133}}{(s-m_1^2)}\right)^2\,\, .\nonumber 
\label{2233}
\eea}
\item PMNS matrix with $\delta=0$
\bea
  U &=&\left( \begin{array}{ccc}
    c_{13}c_{12}                    & s_{12}c_{13} & s_{13}      \\
   -s_{12}c_{23}-s_{23}s_{13}c_{12} &   c_{23}c_{12}-s_{23}s_{13}s_{12} 
                                                  & s_{23}c_{13}  \\
    s_{23}s_{12}-s_{13}c_{23}c_{12} & -s_{23}c_{12}-s_{13}s_{12}c_{23}   
                                                  & c_{23}c_{13}       
                                 \end{array}  \right)\,\, \nonumber
\label{upmns}
\eea  
\item Couplings between different physical scalars obtained from the expression
of potential
{\small
\begin{eqnarray}
-\lambda_{111}&=& 
\lambda_H v_1 a_{11}^3+\lambda_{\Phi} v_2 a_{12}^3
+\lambda_{H\Phi}(v_2a_{11}^2a_{12}+v_1a_{11}a_{12}^2)+\lambda_{HS}
v_1a_{11}a_{13}^2+2\lambda_{\Phi S}v_2a_{12}a_{13}^2
\,\, , \nonumber \\
-\lambda_{222}&=& 
\lambda_H v_1 a_{21}^3+\lambda_{\Phi} v_2 a_{22}^3
+\lambda_{H\Phi}(v_2a_{21}^2a_{22}+v_1a_{21}a_{22}^2)+\lambda_{HS}
v_1a_{21}a_{23}^2+2\lambda_{\Phi S}v_2a_{22}a_{23}^2
\,\, , \nonumber \\
-\lambda_{122}&=&3
\lambda_H v_1a_{11}a_{21}^2+ 3\lambda_{\Phi}v_2a_{12}a_{22}^2
+\lambda_{H\Phi}(v_2(a_{21}^2a_{12}+2a_{11}a_{21}a_{22})+v_1(a_{11}a_{22}
^2+2a_{21}a_{12}a_{22})\nonumber \\ &&
+\lambda_{HS}v_1(a_{11}a_{23}^2+2a_{21}a_{13}a_{23})
+2\lambda_{\Phi S}v_2(a_{12}a_{23}^2+2a_{22}a_{13}a_{23})\,\, , \nonumber \\
-\lambda_{211}&=&3
\lambda_H v_1a_{11}^2a_{21}+ 3\lambda_{\Phi}v_2a_{12}^2a_{22}
+\lambda_{H\Phi}(v_2(a_{11}^2a_{22}+2a_{11}a_{21}a_{12})+v_1(a_{21}a_{12}^2
+2a_{11}a_{12}a_{22})\nonumber \\ &&
+\lambda_{HS}v_1(a_{21}a_{13}^2+2a_{11}a_{13}a_{23})
+2\lambda_{\Phi S}v_2(a_{22}a_{31}^2+2a_{12}a_{13}a_{23})\,\, , \nonumber \\
-\lambda_{133}&=&3
\lambda_H v_1a_{11}a_{31}^2+ 3\lambda_{\Phi}v_2a_{12}a_{32}^2
+\lambda_{H\Phi}(v_2(a_{31}^2a_{12}+2a_{11}a_{31}a_{32})+v_1(a_{11}a_{32}
^2+2a_{31}a_{12}a_{32})\nonumber \\ &&
+\lambda_{HS}v_1(a_{11}a_{33}^2+2a_{31}a_{13}a_{33})
+2\lambda_{\Phi S}v_2(a_{12}a_{33}^2+2a_{32}a_{13}a_{33})\,\, , \nonumber \\
-\lambda_{233}&=&
3\lambda_H v_1a_{21}a_{31}^2+ 3\lambda_{\Phi}v_2a_{22}a_{32}^2
+\lambda_{H\Phi}(v_2(a_{31}^2a_{22}+2a_{21}a_{31}a_{32})+v_1(a_{21}a_{32}
^2+2a_{31}a_{22}a_{32})\nonumber \\ &&
+\lambda_{HS}v_1(a_{21}a_{33}^2+2a_{31}a_{23}a_{33})
+2\lambda_{\Phi S}v_2(a_{22}a_{33}^2+2a_{32}a_{23}a_{33})\,\, , \nonumber \\
-\lambda_{1133} &=&
\frac{3}{2}(\lambda_{H}a_{11}^2a_{31}^2)+\frac{3}{2}(\lambda_{\Phi}a_{12}^2a_{32
} ^2)+\frac{3}{2}(\lambda_{S}a_{13}^2a_{33}^2)
+\frac{\lambda_{H\Phi}}{2}(a_{12}^2a_{31}^2+a_{11}^2a_{32}^2+4a_{11}a_{12}a_{31}
a_{32})\nonumber \\&&
+\frac{\lambda_{HS}}{2}(a_{11}^2a_{33}^2+a_{13}^2a_{31}^2+4a_{11}a_{13}a_{31}a_{
33})
+\lambda_{\Phi S}(a_{12}^2a_{33}^2+a_{13}^2a_{32}^2+4a_{12}a_{13}a_{32}a_{33})
\,\, , \nonumber \\
-\lambda_{2233} &=&
\frac{3}{2}(\lambda_{H}a_{21}^2a_{31}^2)+\frac{3}{2}(\lambda_{\Phi}a_{22}^2a_{32
} ^2)+\frac{3}{2}(\lambda_{S}a_{23}^2a_{33}^2)
+\frac{\lambda_{H\Phi}}{2}(a_{22}^2a_{31}^2+a_{21}^2a_{32}^2+4a_{21}a_{22}a_{31}
a_{32})\nonumber \\&&
+\frac{\lambda_{HS}}{2}(a_{21}^2a_{33}^2+a_{23}^2a_{31}^2+4a_{21}a_{23}a_{31}a_{
33})
+\lambda_{\Phi S}(a_{22}^2a_{33}^2+a_{23}^2a_{32}^2+4a_{22}a_{23}a_{32}a_{33})
\,\, , \nonumber \\
-\lambda_{3333}&=&
\frac{1}{4}(\lambda_{h}a_{31}^4+\lambda_{\Phi}a_{32}^4+\lambda_{S}a_{33}^4)
+\frac{\lambda_{H\Phi}}{2}a_{31}^2a_{32}^2
+\frac{\lambda_{HS}}{2}a_{31}^2a_{33}^2
+\lambda_{\Phi S}a_{32}^2a_{33}^2
\,\, . \nonumber
\label{coup}
\end{eqnarray}
}
\end{itemize}

\end{document}